\documentclass[conference]{IEEEtran}
\usepackage{cite}
\usepackage{url}

\ifCLASSINFOpdf
\else
\fi

\usepackage{multirow}
\usepackage{amsmath,amssymb,amsfonts}
\usepackage{caption}
\usepackage{subcaption}

\usepackage{algorithm}
\usepackage[noend]{algpseudocode}
\usepackage{lipsum}
\usepackage{amsthm}
\newtheoremstyle{mystyle}
  {}
  {}
  {\itshape}
  {}
  {\bfseries}
  {.}
  { }
  {}

\theoremstyle{mystyle}
\newtheorem{theorem}{Theorem}
\usepackage{mathtools}
\usepackage{relsize}

\usepackage{algorithm}
\usepackage[noend]{algpseudocode}
\usepackage{graphicx}

\newcommand{\indep}{\rotatebox[origin=c]{90}{$\models$}}
\newcommand{\nop}[1]{}

\begin{document}
\title{CoShare: An Efficient Approach for Redundancy Allocation in NFV}

\author{\IEEEauthorblockN{Yordanos Tibebu Woldeyohannes\IEEEauthorrefmark{1}, Besmir Tola\IEEEauthorrefmark{1}, Yuming Jiang\IEEEauthorrefmark{1}, and K. K. Ramakrishnan\IEEEauthorrefmark{2}}
\IEEEauthorblockA{\IEEEauthorrefmark{1}Norwegian University of Science and Technology, NTNU,
Trondheim, Norway}
\IEEEauthorblockA{\IEEEauthorrefmark{2}University of California Riverside, California, USA}}

\maketitle

\begin{abstract}
An appealing feature of Network Function Virtualization (NFV) is that in an NFV-based network, a network function (NF) instance may be placed at any node. On the one hand this offers great flexibility in allocation of redundant instances, but on the other hand it makes the allocation a unique and difficult challenge. One particular concern is that there is inherent correlation among nodes due to the structure of the network, thus requiring special care in this allocation. To this aim, our novel approach, called \textit{CoShare}, is proposed. 
Firstly, its design takes into consideration the effect of network structural dependency, which might result in the unavailability of nodes of a network after failure of a node. 
Secondly, to efficiently make use of resources, CoShare proposes the idea of \textit{shared reservation}, where multiple flows may be allowed to share the same reserved backup capacity at an NF instance. Furthermore, CoShare factors in the heterogeneity in nodes, NF instances and availability requirements of flows in the design. The results from a number of experiments conducted using realistic network topologies show that the integration of structural dependency allows meeting availability requirements for more flows compared to a baseline approach.   
Specifically, CoShare is able to meet diverse availability requirements in a resource-efficient manner, requiring, e.g., up to 85\% in some studied cases, less resource overbuild than the baseline approach that uses the idea of \textit{dedicated reservation} commonly adopted for redundancy allocation in NFV.
\end{abstract}

\section{Introduction}\label{sec-intr}

Network softwarization is transforming how networks are designed and operated to deliver specialized / innovative services and applications. Network Function Virtualization (NFV) has emerged as the key driver of network softwarization, promising among other aspects, network automation, flexible service provisioning, and cost reduction~\cite{YI2018212}. NFV is also considered as a key enabler for the new generation of communication networks such as 5G cellular networks~\cite{blanco2017technology,CleanGToN},
which support diverse types of services, requiring different levels of availability.
However, the successful adoption of NFV in production networks is associated with new challenges. One of them is to ensure
these diverse availability requirements 
of services provided by an NFV-based network~\cite{mijumbi2016nfv_challenges,tola2019network,Kamisinski2020}. 

The ``de-facto'' technique for an NFV-enabled network to fulfill availability requirements of
its services is through allocation of redundant/ backup network function (NF) resources to compensate for the failures of primary NFs~\cite{tola2019network},   
since the mere provisioning of primary NF service chains is insufficient,  especially for flows requiring carrier-grade services
~\cite{tola2019network,dinh2018efficient,fan2017carrier}. 
The diversity in the availability requirements of flows, implies that the levels of redundancy for them may differ. For example, while one backup chain might suffice to satisfy flows with lower availability requirements, more than one backup chain might be needed to meet higher availability demands.

One of the main factors which needs to be taken into account in redundancy allocation is efficient utilization of resources. In different virtualization technologies considered so far, redundancy is provided in the form of hot-standby replicas of NF instances.
Typical solutions such as VMware Fault Tolerance~\cite{vmware_ha} and the more recent NFV system-level framework~\cite{reinforceTON}
envision the instantiation of a {\em dedicated backup instance}, which runs {\em on a separate node}. 
However, such solutions can be resource inefficient, as each NF requires at least two instances (one primary and one backup). 

Another crucial aspect that needs to be taken into account in redundancy allocation is the effect of the network structure or topology. 
This is due to that, the impact of one node's failure on the services provided by the network may significantly differ from that of another node's failure due to the inherent
dependencies, called {\em network structural dependencies}\cite{Woldeyohannes2018measures, SCGR2019}. For example, failure of node $n$ in Fig. 1 will cause three nodes to be unreachable 
and hence unavailability of all these nodes and their hosted NFs to other nodes in the network. As a result, for a service chain along with a backup chain, if it has an NF whose primary and backup instances were both hosted at these nodes, only one failure could cause both chains to become non-functional.  

In addition, a unique characteristic of NFV is that {\em an NF instance may be hosted at any node in the network}. This appealing characteristic offers additional flexibility in redundancy allocation in NFV, i.e., the ability to decide {\em where to place the backup instances}, in addition to {\em how to assign backup instances} to form backup service chains for flows
\cite{hmaity2017protection,beck2016resilient}. 
 While this allocation problem shares some similarity with the Virtual Network Embedding (VNE) problem \cite{mijumbi2016nfv_challenges,beck2016resilient}, as also highlighted in \cite{beck2016resilient}, specific facets such as allowing multiple flows to share a single virtual NF instance and requiring ordered concatenation of NFs to form a service chain make the problem different from VNE.

Previous works on NFV redundancy allocation consider also the problem of which nodes are assigned for backup instances, but without taking into account the freedom of placing backup instances at different nodes in the network topology~\cite{fan2018framework,fan2017carrier}. More importantly, the problem of jointly addressing the placement and the assignment of service requests to backup instances has not been considered~\cite{kanizo2017optimizing,fan2015grep,herker2015data,ding2017enhancing,kanizo2018designing}. 
Approaches in~\cite{kanizo2017optimizing, he2019optimization} provide node or NF instance level protection that is not tailored to the specific service availability requirements of individual flows. In works such as~\cite{fan2015grep,tomassilli2018resource,qu2018reliability,li2019availability}, the fulfillment of service availability requirements at the flow or NF chain level is considered. 
However, as discussed in the related work section in more detail (Appendix.~\ref{sec-rela}), the restrictive setups and assumptions made by them limit the scope and applicability of those approaches. For example, in~\cite{li2019availability,fan2018framework}, backup instances of a service chain need to be hosted on the same node and in~\cite{fan2018framework,fan2017carrier}, dedicated capacity is reserved for every flow in the assigned backup instances.

In this paper, a novel redundancy allocation approach for NFV-based networks, called CoShare, is proposed. The contributions of CoShare are three-fold. Firstly, CoShare explores the flexibility offered by the unique characteristic of NFV, assigning backup NF instances to nodes meticulously to avoid the potential 
concurrent unavailability of the primary and backup service chains due to correlated failures caused by network structural dependencies, as exemplified above using Fig. 1. For this purpose, the information centrality measure called \textit{node dependency index}~\cite{kenett2012dependency, Woldeyohannes2018measures} is exploited to identify correlation among nodes (see Sec.~\ref{sec:dependency_index}). The correlation information is made use of in both backup instance placement (\S~\ref{sec:place-stru}) and assignment (\S~\ref{sec:assign-stru}). To the best of our knowledge, CoShare is the first redundancy allocation approach for NFV-based networks, which explicitly considers the network structure-caused correlation among nodes in the design. 

Secondly, exploiting the derived correlation information, CoShare proposes to improve resource utilization efficiency by exploiting \textit{shared reservations}, where multiple service chains that are not susceptible to simultaneous service interruption (referred to as independent chains) are allowed to share the same reserved backup capacity of an NF instance. 
To the best of our knowledge, CoShare is the first to adopt this concept, i.e., \textit{shared reservation}, in the NFV context.
In the literature, the general idea of sharing backup resources has long been exploited for backup allocation to improve resource utilization in various types of networks, e.g., MPLS, IP, and optical mesh networks~\cite{Doucette03, li2002efficient,ou2004shared-path,tornatore2005photo,xu2006}. However, there is a fundamental difference. In traditional network settings, the locations of the backup resources are typically fixed in the network, while in an NFV-based network, owing to the flexibility offered by NFV, they cannot or need not be assumed a priori. 
In other words, the placement of backup NF instances can have a significant impact on the decision of how they can be shared, making the problem new and challenging. 
 For NFV redundancy allocation, few works have considered exploiting sharing to improve resource utilization efficiency. Among the approaches described in the literature, either the sharing is at a coarser-grained level, e.g., NFs sharing a host machine ~\cite{qu2018reliability}, or it requires dedicated resources e.g.~\cite{li2019availability}. More discussion is provided in the Related Work section.  

Thirdly, CoShare takes into account the heterogeneity in the availability requirement across flows in allocating back up NF resources, together with the node heterogeneity in supporting NFs, where the heterogeneity in node and instance availability levels is also considered.  

In brief, CoShare places backup NF instances and assigns them to form backup service chains to meet diverse availability requirements of flows by {\em jointly considering the network structural dependency-caused impact} (Sec.~\ref{sec:dependency_index}, Sec.~\ref{sec:place-stru}, Sec.~\ref{sec:assign-stru}), {\em the heterogeneity in nodes, instances and availability requirements} (Sec.~\ref{sec:heterogeneity-1}, Sec.~\ref{sec:heterogeneity-2}), and {\em the efficiency in resource utilization} (Sec.~\ref{sec:efficient}). The main contributions of this paper are summarized as follows:
\begin{itemize}
\item A novel redundancy allocation approach for NFV, called CoShare, is proposed, where an information centrality measure is exploited to identify the inherent correlation among nodes due to the structure of the network (\S~\ref{sec:dependency_index}), based on which backup NF instances are placed (\S~\ref{sec:place-stru}) and assigned  (\S~\ref{sec:assign-stru}) to form the backup NF chains.  

\item The proposed heuristics of CoShare adopts a new concept of \textit{NF shared reservation}, to achieve higher efficiency in resource utilization (\S~\ref{sec:efficient}). Specifically, independent flows with disjoint primary service chains are allowed to share common reserved capacity at an NF instance for fault-tolerance. 

\item The placement and assignment of NF instances takes into account the heterogeneity of nodes, of instances, and required service availability levels, to decide where to place backup NF instances (\S~\ref{sec:heterogeneity-1}) and how to assign NF instances to form backup chains (\S~\ref{sec:heterogeneity-2}).
\end{itemize}

\section{The System Model and Problem Definition}\label{sec-syst}

\begin{figure}[bt]
\centering
  \includegraphics[scale=0.225]{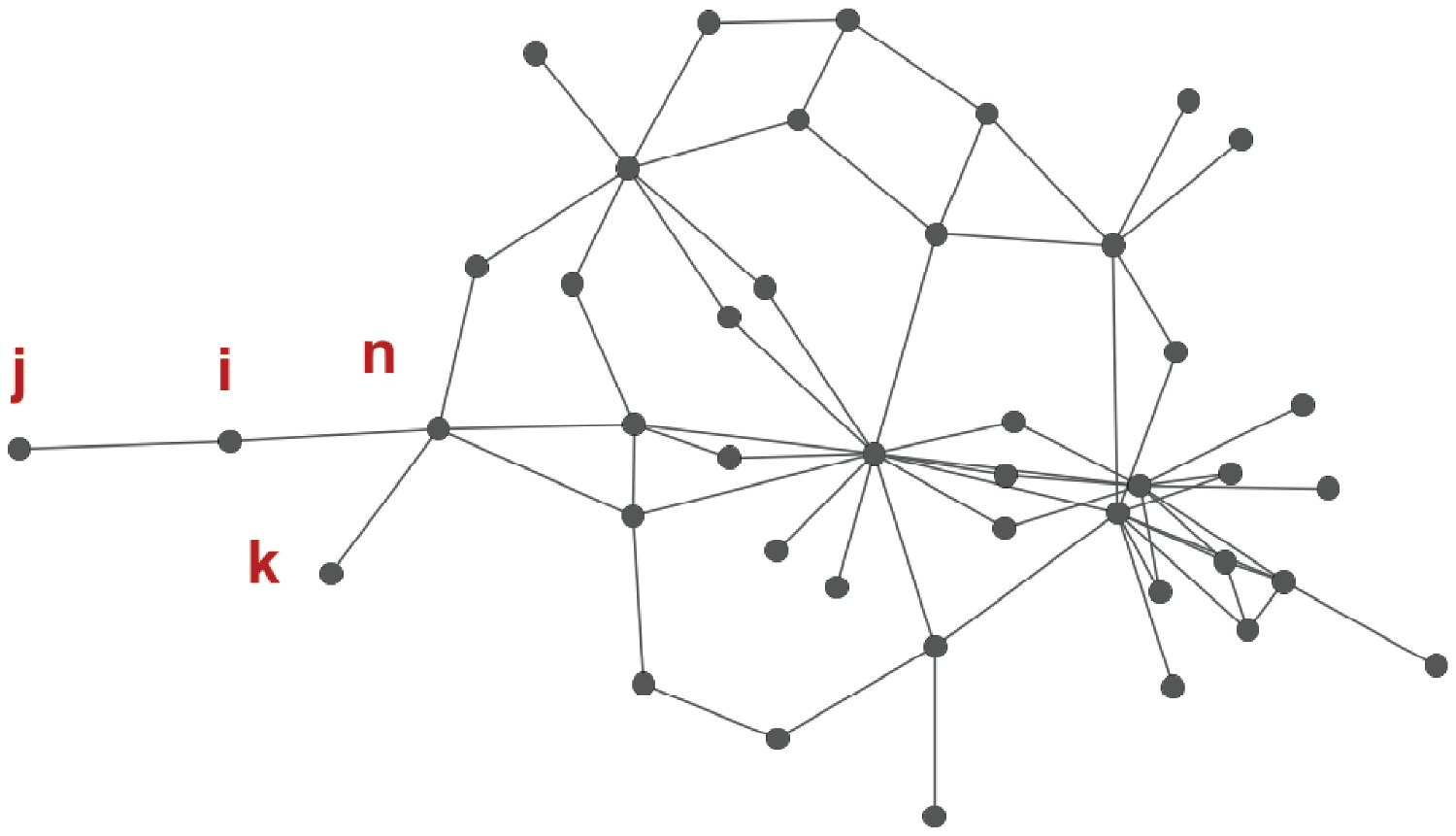} 
 \caption{Network-structurally correlated nodes with node $i$} \label{dependence}
 \vspace{-10pt}
 \end{figure}

\subsection{The Network Model} 

We consider the services provided by an NFV-based network, with the network represented as an undirected connected graph $G(\mathcal{N},\mathcal{L})$, where $\mathcal{N}$ denotes the set of nodes and $\mathcal{L}$ the set of links. 
Each flow $f\in\mathcal{F}$, where $\mathcal{F}$ represents the set of traffic flows requesting service, has a source ($s_f$) and a destination ($d_f$), and has a traffic rate denoted as $\lambda_f$ in packets per second (pps). The flow $f$ is identified by the source-destination pair ($s_f$, $d_f$), where $s_f$ and $d_f$ are in $\mathcal{N}$. We call such source and destination nodes ``end nodes'' in this paper and assume  
that the end nodes are not involved in hosting NF instances. Each of the other nodes may have multiple CPU cores to host NF instances. {\em An NF instance can be hosted on any node in $\mathcal{N}$, which is not a source or destination of a flow, and  has enough available resources}. Keeping with typical deployment approaches \cite{zhang2016opennetvm,woldeyohannes2018cluspr}, we assume that a CPU core, if allocated, is dedicated to a single NF instance. An NF instance may be allocated one or more CPU cores.

The network service provided to flow $f$ is represented by an NF chain $\overrightarrow{S}_{f}$,  i.e., a series of network functions $(S^1_{f}, S^2_{f} \ldots S^{g_f}_{f})$ that are performed in a specified order~\cite{mijumbi2016nfv_challenges}, where $g_f$ denotes the service chain length. 
An NF instance $v$ is assumed to require $k_v$ number of cores and has $\mu_v$ processing capacity (i.e., the amount of traffic the instance can process per unit time). An NF instance may process multiple flows whose service chains include that corresponding NF. It is required that $\sum_{f \in \mathcal{F}^v} \lambda_f \leq \mu_v$, where $\mathcal{F}^v\subseteq \mathcal{F}$ denotes the set of flows processed by $v$\footnote{If there are flows with extremely high traffic rates, NF instances with enough processing capacities should be included, or other mechanisms may be adopted, e.g., splitting the flow into sub-flows, to meet this requirement.}.  

It is assumed that for each flow, its service chain has already been allocated using an NFV resource allocation algorithm, e.g., ClusPR \cite{woldeyohannes2018cluspr}, where, however, the availability aspect of the service has not been taken into account.  
We term this allocated service chain as the {\em primary chain} of the flow.

\subsection{The Availability Model} 

The network service provided to each flow $f$ has an availability requirement $A_{r_f}$. 
For practical reasons, the failure impact of the flow's source and destination nodes is excluded. In addition, to simplify the representation and analysis, we focus on the impact of (hardware) node and (software) NF instance failures on the availability, and assume that nodes and instances fail independently. Reflecting the typical classification of availability requirements, as seen in the specification by the European Telecommunication Standard Institute (ETSI) ~\cite{etsi2015resiliency_req}, we consider three levels of service availability, referred to as classes. We assume that flows fall into: low, medium, and high availability class. As an example, in the evaluation, we assume the availability requirements of flows are in the range of $[99.9\%, 99.99\%)$,  $[99.99\%, 99.999\%)$ and $[99.999\%, 99.9999\%)$ and use them to represent the three classes and also simply refer them as the $99.9\%$, $99.99\%$ and $99.999\%$ classes respectively.  

We are interested in {\em meeting the availability requirements of flows through redundancy}, by allocating additional NF instances. More specifically, if the primary chain cannot provide the required $A_{r_f}$, {\em backup chain(s)} are created to improve the availability level of the service and meet the requirement. The primary chain of flow $f$, denoted with $p_f$, defines a mapping of the network service chain $\overrightarrow{S}_{f}$ with the nodes that host the service chain. Similarly, the backup chain $b_f$ defines a mapping of the same chain to the nodes that will host the backup instances of flow $f$. Note that there can be more than one backup chains, hence mappings, if a single one is not sufficient to meet the flow's availability requirement. The service to a flow is considered available if either the primary or one of the backup service chains is available.

Let $A_n$ denote the availability of node $n$, and $A_v$ the availability of NF instance $v$. For the primary chain of flow $f$, we use $A_{p_{f}}$ to denote its availability. For a backup chain $b_f$, its availability, denoted as $A_{b_f}$, is calculated as: 
\begin{equation}\label{availeq01}
A_{b_f}= \prod_{g=1}^{g_f} A_{v_g}\prod_{n \in \mathcal{N}_{b_f}} A_{n}
\end{equation}
where $\mathcal{N}_{b_f} (\subset \mathcal{N})$ denotes the set of nodes that host NF instances of the backup chain $b_f$, and $v_g$ the instance on $b_f$ for the $g$-th NF of the service chain. 
If the backup chain(s) and the primary chain do not share a host node, the overall service availability $A_f$ is given by a parallel combination: 
\begin{equation}\label{availeq02}
A_f = 1-(1-A_{p_f})\prod_{b_f \in B_f} (1-A_{b_f})
\end{equation}
where $B_f=\{b_f^1,\dots,b_f^{h_f}\}$ denotes the set of backup chains and $h_f=|B_f|$ denotes the set cardinality.

\subsection{The Redundancy Allocation Problem}\label{sec-d123} 
For an NFV-based network, the redundancy allocation problem has three aspects to consider, which are, 
\begin{itemize}
\item \textit{(C1) Deciding the number of backup instance for each NF:} 
The needed number for each NF depends on the availability requirements of flows, the availability and capacity characteristics of both the nodes and the NF instances, as well as the topology of the network. When the availability requirements are stringent while nodes and/or NF instances have low availability values, more than one backup chain may have to be allocated \cite{fan2017carrier,zhang2019raba}. 
\item \textit{(C2) Placing the backup instances:} 
Similarly, the placement of backup instances is also influenced by the same factors. For example, the placement of a backup instance should not only comply to anti-affinity constraints~\cite{isg2016network_resiliency} to avoid common failure modes, but it should also take into account the node's resource capabilities: placing a backup instance in a given node may happen only if the node has sufficient resources, e.g., CPU cores.
\item \textit{(C3) Assigning instances to form backup chains for flows:}  
This concerns the assignment of instances to each flow to form backup service chains to meet the flow's availability requirement.  Similar considerations apply. For instance, a flow can be assigned backup instances only if those instances have sufficient resource capacity to accommodate the flow.  
\end{itemize}

The three considerations discussed above are entangled, and together with the special NFV characteristic that any node may host instances of any NF, makes redundancy allocation in NFV both unique and challenging. 
To have a better overview of the problem and to achieve efficient resource utilization while providing redundancy allocation, we formulate it as an optimization problem as follows:
\begin{eqnarray}\label{eq-obj}
\text{Given:}  && G(\mathcal{N},\mathcal{L}) \nonumber \\
\text{Minimize}_{\forall (\beta, \alpha)}  && N_v(\beta, \alpha), \qquad \forall v \in \mathcal{V}  \label{opt-problem}
\\
\text{Such that} && A_f(\beta, \alpha) \ge A_{r_f}, \quad  \forall f \nonumber 
\end{eqnarray}
where $\beta$ denotes the adopted NF backup instance placement strategy and $\alpha$ the assignment strategy: Each strategy specifies the values for the related decision variables, such as the number of instances for each NF and the node where an NF instance is placed. 
In addition, $\mathcal{V}$ is the set of NFs involved in providing the services, $N_v(\beta, \alpha)$ denotes the number of instances of NF $v$ and $A_f(\beta, \alpha) $ is the achieved availability for flow $f$, under the strategies $\beta$ and $\alpha$. 

For simplicity, only the topology and the availability conditions / constraints are included in (\ref{opt-problem}). In our earlier work~\cite{woldeyohannes2019towards}, a more complete version of the problem including the constraints can be found. In addition, an Integer Linear Program (ILP) model has been developed in~\cite{woldeyohannes2019towards} to solve the problem.

The optimization problem (\ref{opt-problem}) is similar to the well-known Capacitated Location-Routing problem (C-LRP) that is NP-hard ~\cite{prodhon2014survey,hmaity2016virtual,hmaity2017protection}. Specifically, our problem similarity to C-LRP consists in finding the optimal placement of NFs, among a set of available hosting nodes with different capacities, and finding the optimal paths for routing traffic along the NFs aiming at minimizing the total number of NFs and nodes to be utilized. 
A direct consequence of the problem complexity is that when the network is large, solving the problem optimally in limited time is difficult. For this reason, a heuristic approach, called CoShare, is proposed in this paper to address (C1) -- (C3). The heuristic is introduced in detail in Sec. \ref{sec:heuristic} and Sec. \ref{assignment}. 

It is worth highlighting that {\em redundancy allocation for a backup chain should try to avoid sharing risk of failures with the primary service chain, i.e., the failure of any primary instance or its hosting node should have minimal impact on the backup service chain.}  This is crucial and is also the basis for applying (\ref{availeq01}) and  (\ref{availeq02}). However, even though the nodes and instances may be independent as individual systems, they are inherently correlated due to network structural dependence: the failure of one node may cause the unreachability of other nodes if these nodes have strong network structural dependency with the failed node (Cf. Sec. \ref{sec:dependency_index}). A novel idea of CoShare is to explicitly take into account the inherent correlation due to the network topology in the redundancy allocation problem (\ref{opt-problem}). 

\section{Identifying Correlation among Nodes Due to Network Structural Dependence}
\label{sec:dependency_index}

\subsection{Network Structural Dependency Measure}
The inherent structural dependencies among nodes imply that the impact of one node's failure on the services provided by the network may significantly differ from the failure of another node. To reflect this difference, several measures have been proposed in the literature~\cite{kenett2012dependency,Woldeyohannes2018measures}. 
For the NFV redundancy allocation problem, a key is to choose the proper nodes to place the backup instances. To this aim, the \textit{node dependency index} \cite{Woldeyohannes2018measures, kenett2012dependency} is adopted. 

The \textit{node dependency index} $DI(i|n)$ measures the average level of dependency that node $i$ has on node $n$ in connecting to the other nodes of the network~\cite{Woldeyohannes2018measures}. Specifically, $DI(i|n)$ is calculated from the path dependency index $DI(i\rightarrow j|n)$, which measures the dependency that the path between nodes $i$ and $j$ has on node $n$. $DI(i\rightarrow j|n)$ is defined as 

\begin{equation}\label{linkdpi}
DI(i \rightarrow j | n) \equiv
\begin{cases}
I_{ij}  -  I_{ij}^{-n} &  \text{if} \ {\mathcal{A}}^{-n}_{ij}=1 \\
1 &  \text{if}   \ {\mathcal{A}}^{-n}_{ij}=0,
\end{cases}
\end{equation}

\noindent
where $I_{ij}$ and $I_{ij}^{-n}$ are the information measures between nodes $i$ and $j$ before and after the deactivation of node $n$ \cite{stephenson1989rethinking}. Specifically, they are defined as:
$$
I_{ij} =  1/d_{ij}; \qquad I_{ij}^{-n} =  1/d_{ij}^{-n}
$$
with $d_{ij}$ and $d_{ij}^{-n}$ denoting, respectively, the length, in terms of hop count, of the shortest path between nodes $i$ and $j$ before and after the disabling of node $n$ \cite{stephenson1989rethinking}. The binary variable ${\mathcal{A}}^{-n}_{ij}$ measures the reachability of node $j$ from node $i$ given that node $n$ has failed: ${\mathcal{A}}^{-n}_{ij}=1$ if node $i$ can reach node $j$ after the deactivation of node $n$ and 0 otherwise. 

The node dependency index is defined as~\cite{Woldeyohannes2018measures} \footnote{In ~\cite{Woldeyohannes2018measures}, $N-1$ is used in the denominator, but since there are only $N-2$ choices for $j \in \mathcal{N}^{-n}/ i \neq j$, the more intuitive $N-2$ is adopted in (\ref{nodedpl}). Note that, this change does not affect the resulting ranking of nodes.}: 
 
 \begin{equation}\label{nodedpl}
DI(i|n) = \frac{1}{N-2}\sum_{j\in \mathcal{N}^{-n} / i \neq j} DI(i \rightarrow j | n).
\end{equation} 
where $N=|\mathcal{N}|$ and $\mathcal{N}^{-n} = \mathcal{N} \setminus \{n\}$ is defined as the set difference between $\mathcal{N}$ and $\{n\}$, i.e., the set of network nodes excluding $n$. 

It can be verified: $0 \le DI(i|n) \le 1$. For the two extreme cases, $DI(i|n)=0$ tells that $i$ does not experience connectivity problem with removal of $n$, while $DI(i|n)=1$ implies that  $i$ is unable to connect with any of the other nodes after $n$'s failure.

Fundamentally, a higher $DI(i|n)$ value indicates higher dependence of $i$ on $n$ due to the network structure. 

{\bf Remark:} Having a longer path with more elements typically reduces the service availability. This motivates us to use (\ref{linkdpi}) as the basis to measure how the reachability between two nodes depends on another node. Essentially,  $DI(i \rightarrow j | n) $ quantifies the extent that this reachability from $i$ to $j$ is affected by the failure of node $n$, and hence the extent that $i \rightarrow j$ depends on $n$. Note that, except for  (\ref{linkdpi}), the other ideas of CoShare do not rely on the specific definition of $DI(i \rightarrow j | n) $, and hence could be readily applicable when other definitions for $DI(i \rightarrow j | n) $ are preferred.

\subsection{Network-Structurally Correlated Nodes } 

As discussed above, even though individual nodes may fail independently, such a failure can affect the communication of other nodes in the network due to the inherent network structural dependence. From the definition of the node dependency index, i.e., (\ref{nodedpl}), if node $i$ has a higher-level dependency on node $n$, the failure of node $n$ will result in greater difficulty for node $i$ to reach the other nodes in the network.

We introduce $\mathcal{C}(i)$ as \textit{the set of critical nodes of node $i$} which node $i$ highly depends on. Node $i$ is said to highly depend on a node $n \in \mathcal{C}(i)$, or in other words, $n$ is critical to $i$, if $DI(i|n)$ is above a given threshold $t_{DI}$: 

\begin{equation}\label{e_critialset}
\mathcal{C}(i)= \{n \mid DI(i|n) > t_{DI}, n \in  \mathcal{N}^{-i} \}
\end{equation}

If $\mathcal{C}(i)$ is empty, it means that $i$ is not highly dependent on the other nodes. For example, in a full mesh network, all nodes are structurally independent of each other as the failure of one node does not affect the connectivity among the others. 

It is worth highlighting that the network structural dependence relation between two nodes, $i$ and $j$, has two directions, i.e., $DI(i|j)$ -- dependence of $i$ on $j$ and $DI(j|i)$ -- that of $j$ on $i$. 
As a consequence, to minimize the influence of network structure-caused correlation, such that node $i$ can be used as a backup for node $j$ and vice versa, we should avoid $j \in \mathcal{C}(i)$ or $i \in \mathcal{C}(j)$. We call this the \textit{first-level dependency} among nodes. 
To elaborate, Fig. \ref{dependence} shows an example, where nodes $n$ and $j$ have first-level dependency with node $i$. In particular, while $n \in \mathcal{C}(i)$ is critical to $i$, $j$ is network-structurally highly dependent on $i$, i.e., $i \in \mathcal{C}(j)$. 

In addition, as easily seen in Fig. \ref{dependence}, a critical node $n$ of $i$ may also be critical to another node $k$ i.e., $n \in \mathcal{C}(i)$ and $n \in \mathcal{C}(k)$. In this case, both nodes $i$ and $k$ depend on the same node $n$. As a result, the failure of such a critical node, e.g. $n$, may result in the unavailability of those nodes that depend on it, e.g., $i$ and $k$, hence presenting a structural correlation that we refer to as the \textit{second-level dependency} among nodes. An implication of this is that: we should avoid using $k$ as a backup for node $i$, even though $k$ is not in $\mathcal{C}(i)$. 

Based on the above analysis, Algorithm \ref{herusicnode} presents the pseudo code of the algorithm for finding the set of nodes, denoted as $\hat{\mathcal{B}}_i$, which are network-structurally correlated with node $i$.  $\hat{\mathcal{B}}_i$ is initially empty, the algorithm starts by finding the set of nodes based on the first-level dependency in both directions (Lines 1-2 and Lines 3-5 respectively). Then, nodes that have the second-level of dependency described above are added (Lines 6-9). 
 
\begin{algorithm}[t]
\caption{Finding network-structurally correlated nodes}  
\textbf{Input:} $G(\mathcal{N}, \mathcal{L})$, $t_{DI}$ \\
\textbf{Output:} $\hat{\mathcal{B}}_i$ 
\begin{algorithmic}[1]
     \State Find $\mathcal{C}(i)$ using (\ref{e_critialset}) 
     \State Insert $\mathcal{C}(i)$ to $\hat{\mathcal{B}}_i$
            
        \For{ $j \in \mathcal{N}^{-i}$}
      \If {$i \in \mathcal{C}(j)$}
        \State Insert $j$ to $\hat{\mathcal{B}}_i$ 
        \EndIf
          \If {$j \in \mathcal{C}(i)$}
          
 \For{ $k \in \mathcal{N}^{-j}$}
         \If {$j \in \mathcal{C}(k)$}
          \State Insert $k$ to $\hat{\mathcal{B}}_i$ 
         \EndIf   
      \EndFor       
            \EndIf 
              \EndFor

 \State return $\hat{\mathcal{B}}_i$
 
\end{algorithmic}
\label{herusicnode}

\end{algorithm}

\section{Placement of Backup NF Instances}
\label{sec:heuristic}

This section focuses on estimating the needed numbers of backup NF instances and deciding where to place them. Originally, we take network structural dependence into account. 
 
\subsection{Estimating Numbers of Backup NF Instances} 
\label{sec:nr_backups}

The number of backup instances for each NF, needed to satisfy the service availability requirements of flows, is influenced by several factors, such as the availability and length of the primary chains and the availability and capacity of the NF instances, in addition to their availability requirements. In addition, flows with higher availability requirements might need more than one backup chain while lower service availability requirements may be fulfilled with only one backup chain~\cite{herker2015data, fan2017carrier,woldeyohannes2019towards}. Further due to the heterogeneity in node availability, instance availability, and service availability requirements of flows, finding the needed numbers of backup NF instances is not trivial. 
In CoShare, we use the following approach to estimate such numbers.

Specifically, we first estimate the number of backup chains needed to fulfill the availability requirement of flows in each class, based on which, the number of needed backup instances is then calculated. For the former, the estimation assumes that each NF instance is hosted at a different node and the backup chains and the primary chain are node disjoined. Then, the number of needed backup chains for a flow in class $c$, denoted as $h_c$, is estimated as: 
\begin{equation}\label{availab_req}
h_c = \min_{ x \in \mathcal{Z}^{+}} \{ x | 1-(1-\min_{f \in \mathcal{F}_c}A_{p_f})(1-\tilde{A}_c^b)^x \geq A_c^r  \}
\end{equation}
with 
$$
\tilde{A}_c^b = (\min_{n \in \mathcal{N}^b}A_n\min_{v \in \mathcal{V}}A_v)^{g} 
$$
where $\mathcal{F}_c$ represents the set of flows in class \textit{c} and $A_c^r$ the {maximum} availability  
requirement for flows in class $c$. $\tilde{A}_c^b$ may be interpreted as the availability of a backup chain for a class $c$ flow, where $\mathcal{N}^b \subset \mathcal{N}$ denotes the set of nodes that have the capacity to host backup NF instances, $\mathcal{V}$ the set of NFs, $A_n$ the availability of a node $n$, $A_v$ the availability of an instance of NF $v$, $g$ the maximum NF chain length of flows in the availability requirement class $c$.  
Since to satisfy the required availability more than one backup chain may be required, $h_c$ is estimated from (\ref{availab_req}), taking into account the parallel effect of these backup chains.

Next, the number of backup instances of each NF $v$, denoted as $z_v$, which are needed to fulfill the service availability requirements of all related flows, is calculated as
\begin{equation} \label{eq-est}
z_v = \sum_{c}z_v(c)
\end{equation}
where $z_v(c)$ denotes the number of backup instances of NF $v$ which are needed for related flows in class $c$ and is simply estimated from, assuming that all flows in the class require $h_c$ number of backup chains:
\begin{equation} \label{eq-est2}
z_v(c) =h_c \lceil \frac{\sum_{f \in \mathcal{F}_c/v \in \overrightarrow{S}_{f}}\lambda_f}{\mu_v} \rceil
\end{equation}

Note that, the estimation (\ref{availab_req}) has adopted conservative assumptions, e.g. NFs of a chain are hosted at different nodes, to get (\ref{availab_req}).  In comparison with the literature approaches in \cite{fan2015grep,tomassilli2018resource,qu2018reliability,li2019availability}, they may instead assume instances for the same NF chain to be hosted at the same backup node. 
In addition, with the estimation (\ref{eq-est}), it can be expected that the estimated numbers of backup instances are higher than the optimal numbers needed to fulfill the availability requirements of flows. However, we highlight that, the numbers from (\ref{eq-est}) are only ``rough'' estimates as the starting point. By setting the objective in assigning them to form backup chains to be maximizing the utilization of the backup instances (cf, Sec.~\ref{assignment}), the actually used and hence needed numbers of backup instances can be significantly reduced (cf, Sec.~\ref{sec-bu-no}).

\subsection{Placement of the Backup Instances}\label{sec:place-stru}\label{sec:heterogeneity-1}

After the numbers of backup instances are estimated, CoShare places them on nodes. The heuristic is presented in Algorithm~\ref{placement}. The placement is made by performing \textit{bin-packing}~\cite{martello1990bin} of the NF instances on the nodes, where the \textit{network structural correlation} among nodes, the \textit{heterogeneity} in the availability level of nodes and NF instances, and in the availability requirements of flows, and the number of backup instances for each NF, are particularly taken into consideration.

On the node side for bin-packing, nodes are categorized and prioritized. Specifically, to avoid simultaneous unavailability of the primary and backup chains, it is intuitive that the NFs of a backup chain for a flow should avoid being hosted on those nodes that are ``critically'' correlated with the nodes hosting the primary NF chain of the flow. To this aim, based on the structural correlation information input, $\hat{\mathcal{B}}_n$, from Algorithm 1, nodes are categorized into two sets, $Q_{n}'(c)$ and  $Q_{n}''(c)$, where the former represents the set of nodes that are not structure-critically correlated with the primary nodes of flows in class $c$, and the latter the rest. In the placement or bin-packing, as the intuition indicates, nodes in $Q_{n}''(c)$ are considered only after nodes in $Q_{n}'(c)$ have been exhausted (Lines 5 - 8).

In addition, nodes may have different availability levels, e.g., high-end nodes having 99.9\% availability or higher while low-end nodes having 99\% availability or lower. In~\cite{woldeyohannes2019towards}, it is shown that for the low availability requirement class, it is more cost-efficient to use the low-end nodes. While for the medium and high availability requirement classes, using high-end nodes is preferable as this will lead to the use of fewer backup instances and nodes. Considering these, CoShare prioritizes nodes based on their availability levels. This prioritization is translated into the sorting of nodes (Lines 3 and 4), before the bin-packing is performed.

On the instance side for bin-packing, prioritization is also performed.  
Specifically, NF type that has greater number of backup instances to be placed is given higher priority to be placed.
The underlying intuition is, the estimate (\ref{eq-est2}) implies that more flows require this NF and its backup instances to achieve their availability requirements. Hence, giving it priority will more likely accommodate a higher number of flows~\cite{woldeyohannes2018cluspr}. This prioritization is reflected in sorting the NF instances based on their numbers (Line 1). 

\begin{algorithm}[t]
\caption{CoShare's Placement Heuristic}\label{placement}
\footnotesize{
\textbf{Definitions:}\\
$\mathcal{N}^{c} \gets$ set of nodes hosting primary instances of class $c$ flows\\
$Q_{n}'(c) \gets$ priority queue of structurally uncorrelated candidate backup nodes\\
$Q_{n}''(c) \gets$ priority queue of the other candidate backup nodes, i.e., $\mathcal{N}^{-Q_{n}'(c)}$ \\
$Q_{v}(c) \gets$  priority queue of all NF types $v$ to be placed for each class $c$\\
$z_v(c) \gets$ number of instances of NF type $v$ to be placed for class $c$\\ 
\textbf{Input:} $G(\mathcal{N},\mathcal{L})$, $\hat{\mathcal{B}}_n^{s}= \mathcal{N}^{-\hat{\mathcal{B}}_n}
 $ complement of set $\hat{\mathcal{B}}_n$ (from Algorithm 1) \\
 \textbf{Output:} Placement of backup instances on nodes \\
\textit{Initialize:}\\ 
$ActiveNode \gets \texttt{null}$
\begin{algorithmic}[1] 
 \State $Q_{v}(c)=$ sort $z_v(c)$ in descending order
	\For{each class $c$}
       \State $Q_{n}'(c)$ =  $\bigcap_{n\in \mathcal{N}^{c}} \hat{\mathcal{B}}_n^{s}$ sorted based on node availability
         \State $Q_{n}''(c)$ =  $\mathcal{N}^{-Q_{n}'(c)}$ sorted based on node availability
           \If{$Q_{n}'(c)$ is not empty}
             \State $ActiveNode \gets$ top of $Q_{n}'(c)$
             \Else
              \State $ActiveNode \gets$ top of $Q_{n}''(c)$
             \EndIf
       \While{$Q_{v}(c)$ not empty} 
             \State $v \gets$ NF type from top of $Q_{v}(c)$ with $z_v(c) > 0$
            \While{$Q_{n}'(c)$ and $Q_{n}''(c)$ not empty}
             	\If{$ActiveNode$ has capacity}
            			\State Place $v$ on the $ActiveNode$
            			\State $z_v(c)=z_v(c)-1$ 
            			\State $v \gets$ next NF type from top of $Q_{v}(c)$ 
            	\Else 
            		\State Remove $ActiveNode$ from $Q_{n}'(c)$ or $Q_{n}''(c)$
					 \If{$Q_{n}'(c)$ is not empty}
             \State $ActiveNode \gets$ top of $Q_{n}'(c)$
             \Else
              \State $ActiveNode \gets$ top of $Q_{n}''(c)$
             \EndIf
           		\EndIf 
           \EndWhile                  
      \EndWhile
	\EndFor 
\end{algorithmic}
}
\end{algorithm}

Finally, the placement is completed by bin-packing the backup instances onto the nodes, based on the prioritization introduced above. Specifically the top-prioritized node is checked for its capacity. If it has enough capacity, the NF instance with the highest priority is placed on it. If not, the next prioritized node is checked for possible placement of this instance (Lines 9-21). As a highlight, a node may have capacity to host multiple NF instances. In such a case, CoShare diversifies the types of instances placed on the node. The underlying intuition is that, the backup chain delay is minimized if all its NFs are hosted on the same node, and placing instances of different types on one node increases this chance. This is reflected by Lines 13-15 in Algorithm 2, where after placing a given NF type on a node, the next NF type from queue $Q_{v}(c)$ is chosen to be placed on the same node instead of another instance of the same type. This procedure is repeated until all the instances are placed or all the backup resources are utilized.

\section{Assignment of NF Instances to Flows} \label{assignment}

The goal of redundancy allocation in NFV is to achieve the desired availability levels for flows. To this aim, having estimated the needed backup instances for each NF and decided where to place them in Sec. \ref{sec:heuristic}, CoShare assigns NF instances to flows to form backup chains for them so as to meet their availability requirements, which is the focus of this section. Some literature works also refer this problem as the flow routing problem or service chaining problem~\cite{hmaity2016virtual,CASAZZA201947}.

\subsection{Feasible NF Instance Set for Assignment}\label{sec:assign-stru}\label{sec:heterogeneity-2} 

Note that every backup service chain of a flow must include all the NFs ordered in the same way as the primary chain. However, for every NF, it may have multiple instances. This subsection is devoted to identifying a set of such instances, called the ``feasible set'', which are considered for the assignment in CoShare. By {\em feasible set}, it is meant that with  the backup chain formed by any combination of related instances in the set, the flow's availability requirement can be met. 

It is obvious that any instance without enough capacity to accommodate the flow should not be included in the feasible set. In addition, as discussed in Sec. \ref{sec:dependency_index}, an NF instance whose hosting node has critical structural correlation with a node of the primary chain (C.f. Algorithm 1) should not be included in the feasible set either. Let $\tilde{\mathcal{I}}_{S_{f}^g}$, $g=1, \dots, g_f$ denote the resultant instance set of each NF $S_{f}^g$ of the flow.

Since each instance and the hosting node have implicit availability levels, this information can be made use of to find the feasible set.  
In particular, for one backup chain, the best availability range, which can be achieved, is between $(MIN, MAX)$ which are calculated as

\begin{eqnarray}
MIN &=& \min_{v_g \in \tilde{\mathcal{I}}_{S_{f}^{g}}, \forall g, \forall b} 1 - (1 - A_{p_f})(1 - A_{b_f}) \label{aeq1}
\\ 
MAX &=& \max_{v_g \in \tilde{\mathcal{I}}_{S_{f}^{g}}, \forall g, \forall b} 1 - (1 - A_{p_f})(1 - A_{b_f}) \label{aeq2}
\end{eqnarray}
where $A_{b_f}$ is found from (\ref{availeq01}). Note that (\ref{availeq01}) has two parts: $\prod_{g=1}^{g_f} A_{v_g}$ that is the availability resulted from the involved instances $v_g$ and $\prod_{n=1}^{|\mathcal{N}_{b_f}|} A_{n}$ that is the availability resulted from the involved hosting nodes $\mathcal{N}_{b_f}$. When all instances $v_g$ are hosted at different nodes, Eq. (\ref{availeq01}) can be re-written as 
$$
A_{b_f}= \prod_{g=1}^{g_f} A_{v_g} A_{n(v_g)}
$$
where $n(v_g)$ denotes the node hosting the instance $v_g$. 
Intuitively, when the required availability $A_{r_f}$ is smaller than $MIN$, this implies that all possible combinations out of $\tilde{\mathcal{I}}_{S_{f}^g}$ to form a backup NF chain for the flow are able to help meet the requirement and hence the set is already a feasible set. When $MIN < A_{r_f} < MAX$, it means the required availability can be achieved by some (but not all) combinations of instances in $\tilde{\mathcal{I}}_{S_f^g}$. In other words, this set is not a feasible set yet, and additional effort is needed as explained below. 

\begin{figure}[t]
\centering
  \includegraphics[width=0.9\columnwidth]{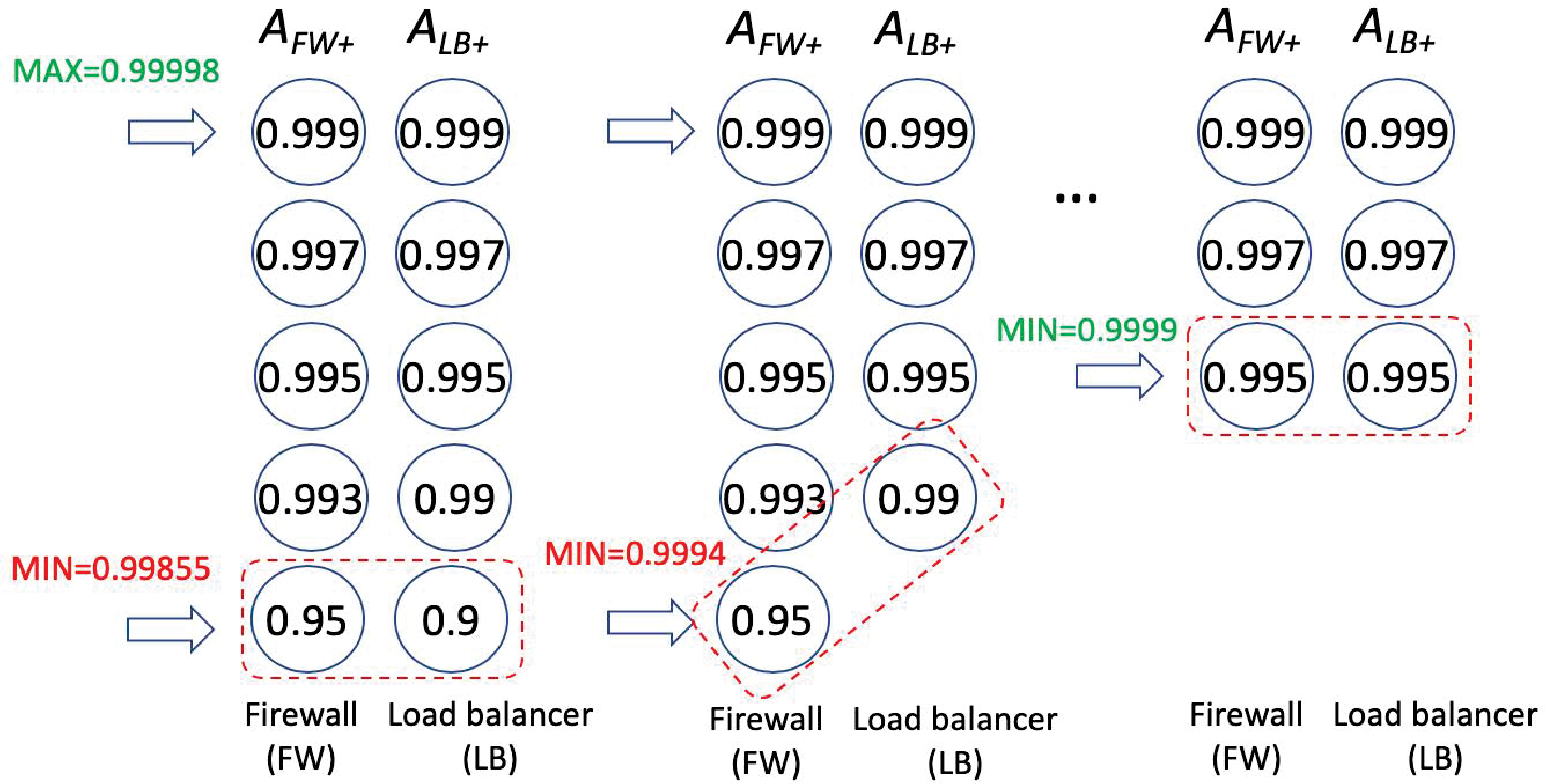} 
\caption{Example of finding feasible instance set (Availability requirement: 0.9999; Availability of the primary chain: 0.99)} \label{feasibleset}
\vspace{-10pt}
\end{figure}

CoShare uses Algorithm \ref{f-set} to find the feasible set. To illustrate the idea, Fig.~\ref{feasibleset} shows a simple example. 
In the example, the considered flow needs a service composed of two NFs, firewall and load-balancer, and its availability requirement is 0.9999, but the primary service chain only has availability of 0.99. In the network, after excluding those instances without enough capacity left or whose nodes are network-structurally correlated with the primary chain nodes, there are still five candidate instances of each NF hosted at different nodes. After sorting, their availability levels, $A_{FW+} = A_{FW}\cdot A_n$ and $A_{LB+} = A_{LB}\cdot A_{n'}$, taking into account both node availability and instance availability, are shown in the figure. 

Clearly, the required availability level 0.9999 is within this $(MIN, MAX)$. An implication is that this requirement can be met with one backup chain. Another is that, some of the instances should not be included in the feasible set. To this aim, we drop the instance with the lowest availability, which is the load balancer instance with availability of $0.9$, and then re-calculate $MIN$ as shown in the middle part of Fig.~\ref{feasibleset} and compare it with the required availability. This process is repeated until $MIN$ is equal to or higher than the required availability. All the remaining instances form the feasible set. 

In the above discussion and example, the required availability level is less than $MAX$. If however this is not true, it means that one backup chain is not enough to fulfill the service availability requirement for the flow. In such cases, we update $A_{p_f}$ with $MAX$ in Algorithm \ref{f-set} when applying (\ref{aeq1}) and (\ref{aeq2}), add the corresponding instances to the assignment and remove them from the candidate lists, and consider using an additional backup chain. This process is repeated until the required availability can be achieved, or there are not enough candidate instances left to form additional backup chains, implying the availability requirement is infeasible to achieve.

\begin{algorithm}[t]
\caption{Finding feasible set of NF instances}\label{f-set}
\footnotesize{
\textbf{Definitions:}\\
${\mathcal{K}_v} \gets$ set of NF instances of type $v$ {with capacity for processing $f$} \\
$A_{v+} \gets A_v*A_b$, where $b$ is the backup host node of NF $v \in \mathcal{V}$\\
\textbf{Output:} set of feasible NF instances $\{ \mathcal{I}_{S_f^1}, \cdots, \mathcal{I}_{S_f^{g_f}}\}$ \\
\begin{algorithmic}[1]
\If {${\mathcal{K}_v}\neq \emptyset$}
 \State Sort the instances in $\mathcal{K}_v$ in descending order of $A_{v+}$  
 \State Find $MIN$ from the instances that have $\min A_{v+}$ using Eq.~(\ref{aeq1}) 
 \State Find $MAX$ from the instances that have $\max A_{v+}$ using Eq.~(\ref{aeq2}) 
  
\If {$MIN \geq A_{r_f} $ $\&$ $|\mathcal{N}_{b_f}|=g_f $}
    \State Insert the instances to the feasible set 
\Else
    \While{$(MIN < A_{r_f} {\leq MAX})$ $||$ $(|\mathcal{N}_{b_f}|<g_f )$} 
        \If {$MIN \geq A_{r_f}$}
            \State Insert the instances making $\min A_{v+}$ to the feasible set
        \EndIf  
        \State Replace the instance with the smallest availability
        \State Recalculate $MIN$        
    \EndWhile 
    \State Insert the instances to the feasible set 
\EndIf
\Else
    \State {Reject $f$}
\EndIf
\end{algorithmic}
}
\end{algorithm}

\subsection{Feasible Backup Chains}
\label{sec:delay_min}

Let $\{ \mathcal{I}_{S_f^1}, \cdots, \mathcal{I}_{S_f^{g_f}}\}$ denote the feasible NF instance set, where $\mathcal{I}_{S_f^v}$, $v = 1, \dots, g_f$, represents the set of instances of network function $S^v_f$ in the feasible set. Then, the possible backup chains for the flow are easily obtained as: 
$$
(I_{S_f^1}, \cdots, I_{S_f^{g_f}}) \equiv \mathcal{R}(f) \quad\quad \forall I_{S_f^v} \in \mathcal{I}_{S_f^v}
$$
 where $v = 1, \dots, g_f$ and $I_{S_f^v}$ denotes an instance of NF $S^v_f$. It is easily verified that, the total number of such possible backup chains is:
$$
|\mathcal{I}_{S_f^1}| \cdots |\mathcal{I}_{S_f^{g_f}}|.
$$

\subsection{Assignment of NF Instances to Flows}
\label{sec:efficient}
In this subsection, we introduce the assignment strategy of CoShare. In brief, for each flow $f$, out of the feasible backup chains, CoShare assigns to the flow the chain that maximizes the utilization of resources so as to minimize the total number of backup NF instances required.  

\subsubsection{Backup capacity reservation}

Since each flow $f$ has an arrival rate $\lambda_f$, every backup NF instance assigned to the flow needs to also reserve $\lambda_f$ amount of capacity to the flow. 
As a consequence, it is intuitive to reserve the same amount of resource for every backup chain as for the primary chain, where dedicated capacity is reserved at every instance~\cite{fan2018framework,fan2015grep,fan2017carrier}. We call this approach {\em dedicated reservation}. Here, it is worth highlighting that, even in this approach, there is sharing at the instance level, i.e., the capacity of the instance can be shared by backup chains of multiple flows as long as the capacity constraint allows, i.e. $\sum_{f \in \mathcal{F}^v} \lambda_f \leq \mu_v$.

However, in practice, the probability that multiple independent failures occur at the same time is low and planning the redundancy considering this rare occasion is costly in terms of resource utilization~\cite{bejerano2005algorithms}. Taking this into consideration, the idea of sharing reserved backup capacity among multiple flows has long been exploited in backup allocation to improve resource utilization in different types of networks~\cite{li2002efficient,ou2004shared-path,tornatore2005photo,xu2006}. In CoShare, we propose to adopt the same idea to reduce the needed numbers of backup NF instances in redundancy allocation. We call this approach {\em shared reservation}. 

\begin{figure}[t]
\centering
\includegraphics[width=0.45\columnwidth]{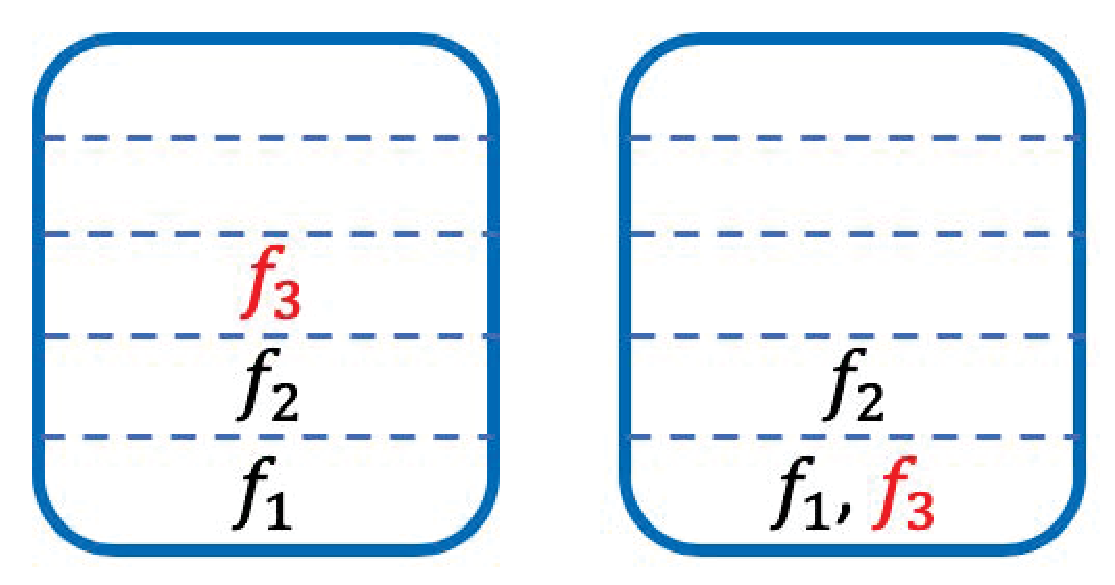}
 \caption{Illustration of dedicated reservation (left) v.s. shared reservation (right): The instance has capacity of 5. Each flow requires capacity of 1. Flows $f_1$ and $f_3$ are independent.} 
 \label{sharedvsdedicated}
\vspace{-10pt}
\end{figure}

Specifically, in CoShare, flows with disjoint primary service chains, referred to as {\em independent flows}, are allowed to share reserved capacity at a backup instance. 
Let $\mathcal{F}_v$ denote the set of independent flows whose service chains use backup instance $v$. Then, in CoShare with shared reservation, the backup NF instance $v$ will only need to reserve a capacity of $\max_{f \in \mathcal{F}_v} \lambda_f$ for all these flows.  In comparison, if dedicated reservation is used, for the same set of flows $\mathcal{F}_v$, the backup instance will need to reserve a capacity of $\sum_{f \in \mathcal{F}_v} \lambda_f$ for them. In Fig.~\ref{sharedvsdedicated}, an example illustrating the difference in capacity allocation between  shared reservation and dedicated reservation is presented.

\subsubsection{The assignment heuristic}
Note that for each flow, a set of feasible backup chains, denoted as $\mathcal{R}(f)$, which together with the primary service chain can meet the availability requirement of the flow has been identified in the previous subsection. Now, the challenge is how to apply the idea of shared reservation in this special setup, i.e. to choose / assign  backup chain(s) to the flow to meet Objective (\ref{eq-obj}). 
To this aim, CoShare utilizes a weight-based approach: 

For each chain $r \in \mathcal{R}(f)$, it is given a weight $W^p(r,f)$ calculated as
\begin{equation}\label{eq-pw}
W(r,f)= \sum_{g =1}^{g_f} w(v^r_{f,g})
\end{equation}
which is the summation of the weight of each instance constituting $r$, where $w(v^r_{f,g})$ denotes the weight of the instance of the $g$-th NF in $r$ for flow $f$, $v^r_{f,g}$ this instance, and $g_f$ the service chain length for flow $f$.

In (\ref{eq-pw}), the instance weight $w(v^r_{f,g})$ is calculated using:
\begin{equation}\label{instw}
w(v^r_{f,g})=
  \begin{cases}
    g_f, & f \indep f_a \text{\&}  f \indep \mathcal{SR}_{v^r_{f,g}}(f_a),  f_a \in \mathcal{D}_{v^r_{f,j}}\\ 
   \frac{\sum_{f_a \in \mathcal{D}_{v^r_{f,g}}}\lambda_{f_a}}{\mu_{v^r_{f,g}}}, & \text{otherwise}.
  \end{cases}
\end{equation}
\nop{
\begin{equation}\label{instw}
\resizebox{\columnwidth}{!}{
  $w(v^r_{f,g})=$
  \begin{cases}
    g_f, & f \indep f_a \quad \text{\&} \quad f \indep \mathcal{SR}_{v^r_{f,g}}(f_a), \quad  f_a \in \mathcal{D}_{v^r_{f,j}}\\ 
   \frac{\sum_{f_a \in \mathcal{D}_{v^r_{f,g}}}\lambda_{f_a}}{\mu_{v^r_{f,g}}}, & \text{otherwise}.
  \end{cases}
  }
\end{equation}
}
where $\mathcal{D}_{v}$ denotes the set of flows that are reserving backup capacity at instance $v$, $\mathcal{SR}_{v}(f)$ the set of flows that are sharing the capacity of instance $v$ with flow $f$, $\lambda_{f}$ the rate of flow $f$, and $\mu_v$ the capacity of instance $v$. In addition, in (\ref{instw}), $\indep$ is used to represent independence between flows. 

The key idea of (\ref{instw}) is as follows. If flow $f$ is independent with flow $f_a$, i.e. $f \indep f_a$, as well as all flows that are sharing capacity of the instance $v^r_{f,g}$ with flow $f_a$, i.e., $f \indep \mathcal{SR}_{v^r_{f,g}}(f_a)$, then it implies that flow $f$ can {\em  share reserved capacity} with flow $f_a$ at the instance $v^r_{f,g}$. In this case, the instance will be given a weight equal to the service chain length of the flow, i.e. $g_f$. Otherwise, the instance is given a weight that is equal to the current utilization level of the instance.

In $\mathcal{R}(f)$, 
the backup chain that has the maximum chain weight i.e., $b_f=\max_{r \in \mathcal{R}(f)} W(r,f)$ is assigned for the flow $f$. 
This assignment strategy prioritizes the assignment of flows to instances that are utilized more, as implied by (\ref{instw}), so as to  minimize the number of instances of each NF needed in Objective (\ref{eq-obj}).

\subsection{Efficiency and Scalability of CoShare}
Theorem \ref{outweigh} ensures that the assignment heuristic of CoShare gives priority to feasible backup chains where shared reservation is possible, thus achieving higher resource efficiency.

\begin{theorem}\label{outweigh}
Consider two backup chain choices $r$ and $r'$ $(\in \mathcal{R}(f))$. The other conditions are the same, but $r$ has at least one instance $v^r_{f,g}$  on which the flow can share reserved capacity with other flows, while $r'$ does not have. Then, $r$ has a larger chain weight than $r'$, i.e., 
$$W(r,f) > W(r^{'},f).$$
\end{theorem}

The complexity of CoShare is determined by the three involved parts, namely network structural analysis (Sec. \ref{sec:dependency_index}), placement (Sec. \ref{sec:heuristic}), and  assignment (Sec. \ref{assignment}). 
In brief, the complexity of CoShare can be written as a function of the number of flows that are to be assigned backup chains, the number of nodes / instances in the network, and the longest length of service chains, which is summarized as Theorem \ref{complexity}. 
\begin{theorem}\label{complexity}
 CoShare has a complexity of $O(FN^G)$, where $F$ is the number of flows, $G$ the longest length of service chains and $N$ the number of nodes in the network. 
\end{theorem}

The proofs of Theorems 1 and 2 are  in Appendix~A.

\section{Results and Discussion}\label{sec-resu} 

This section presents results showing the performance of the proposed redundancy allocation approach, CoShare. Recall that, a novel idea of CoShare is to exploit network structural correlation information in the design. Sec. \ref{sec-6.1} is hence devoted to showing the impact of such correlation. The remaining Sec. \ref{sec-6.2} -- Sec. \ref{sec-6.5} focus on introducing the performance of CoShare where comparisons are also included. 

Specifically, a number of experiments are conducted on two realistic ISP network topologies. In the study, if not otherwise specified, it is assumed that each node hosts 8 CPU cores and has 16 GB memory, of which half of the capacity i.e., 4 CPU cores and 8 GB of memory, is used by the primary NFs and the rest by backup NFs. The primary NFs' placement as well as the assignment of primary service chains to flows is conducted by using ClusPR~\cite{woldeyohannes2018cluspr}. 
Every NF instance requires one CPU core and 2GB of memory, having a total NF processing capacity ($\mu_v$) of 10Mpps. 

Five types of NFs are considered (e.g., Firewall, DPI, NAT, IDS, and Proxy). 
The availability requirements of flows are set according to three levels, namely $99.9\%$ (3'9s), $99.99\%$  (4'9s) and $99.999\%$ (5'9s). 
The NF processing capacity required by each flow $f$, i.e., $\lambda_f$, is set to 0.5 Mpps.
The length of the service chain for each flow is assumed to vary in the range of 2 to 4. 
NFs and the service types in the chain of each flow are selected randomly out of the five NFs considered. 
The availability of the hosting nodes is assumed to be uniformly distributed between $0.99 - 0.999$,  NF instances have an availability between $0.999 - 0.9999$, and the threshold algorithmic parameter $t_{DI}$ is set to 0.5 if not otherwise specified. If a flow's availability requirement cannot be satisfied by the adopted redundancy allocation approach, it is rejected. All algorithms and simulations are run on a Dell workstation with a single Intel$^{\tiny{\textregistered}}$ Xeon$^{\tiny{\textregistered}}$ octa-core processor with 2.4 GHz base frequency and 64 GB of RAM. The results shown are average values with 95\% confidence intervals over ten simulation runs for each availability level.

\subsection{Impact of Network Structural Correlation} \label{sec-6.1}
In this subsection, a simple experiment is carried out to showcase the effect of not considering the network structural correlation among nodes in the backup instance placement decision making. Two different network topologies are used, the GEANT network~\cite{geant:2019} and a simple Barab\`asi-Albert (BA) scale-free network~\cite{barabasi2003scale}. The GEANT network is a pan-European network connecting research and education institutions consisting of 44 nodes and 136 links. (To visualize this, a version with 44 nodes and 68 links is shown in Fig.~\ref{dependence}.) The BA-network node linkage follows a power-law distribution and consists of 45 nodes and 45 links generated with an initial seed of $m_0=5$ nodes. Only node availability impact is considered and it is assumed that the availability of each node is 0.999 (3'9s). For each flow, the service function chain contains two NFs, and the required availability of the flow is $99.999\%$. The baseline algorithm from~\cite{fan2017carrier} is used to decide the number of backup instances needed. According to the baseline algorithm~\cite{fan2017carrier}, theoretically, one backup for each of the NFs is enough to meet the $99.999\%$ (5'9s) availability requirement. 

Two backup instance placement strategies are considered. One does not consider network structural correlation, where the primary and backup NF host nodes of a chain are randomly chosen in the network. Another takes the structural correlation into consideration, where the backup host nodes are chosen from those without critical network structural correlation with the primary host nodes. 

\begin{figure}[bt]
\centering
  \includegraphics[scale=0.45]{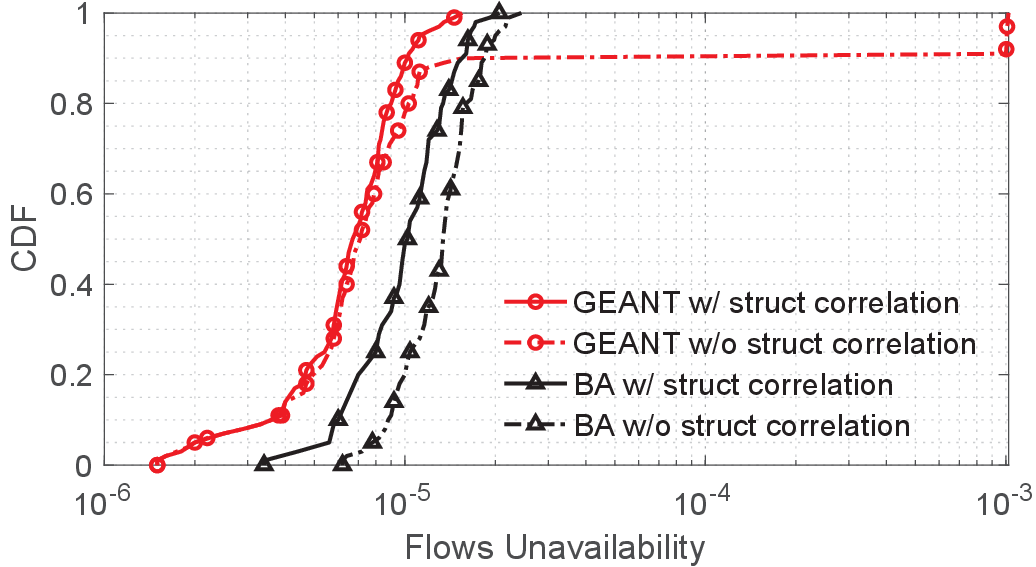} 
 \caption{Impact of Network Structural Correlation}
\label{randavail}
\vspace{-10pt}
\end{figure}

The availabilities of 100 flows are measured by conducting ten million simulation runs. In each simulation run, the state of each node, i.e., failed (0) or up (1), is randomly generated from Bernoulli distribution using the node's availability. 
The unavailability CDF of the 100 flows is shown in Fig.~\ref{randavail}. As can be seen from the figure, the strategy taking network structural correlation into consideration performs significantly better for both networks. Specifically, when network structural correlation is not considered, about $10\%$ of the flows only have availability of 3'9s or even lower for the GEANT network, in contrast to taking it into consideration where all flows have at least 4'9s availability. In addition, while only $80\%$ of flows can reach the required 5'9s in the former, this percentage increases to $90\%$ in the latter.  Moreover, the effect of factoring in the structural correlation is more pronounced for the BA network scenario. All the flows achieve higher availability and 30\% more flows are able to reach the target availability requirement of 5'9s compared to a placement strategy that does not consider structural correlation. The rationale behind this is due to the preferential attachment of BA networks, which tend to create highly connected nodes, also called hubs, whose failure can fragment the network into small isolated islands. From a robustness perspective, this property represents the Achilles' heel of scale-free networks~\cite{barabasi2003scale}. As a result of such fragmentation, the reachability of nodes belonging to separate islands will be interrupted, hence an effective placement of redundancy needs to consider such effects. The ability of the structural correlation in capturing the reachability of nodes upon failure of others helps steer the backup placement more effectively.

There are two implications of this experimental study. One is that the number calculated by the baseline algorithm  \cite{fan2017carrier} may not be enough to meet the availability requirements of all flows when applied to a real network. Another is that, the inherent network structural correlation among nodes can have significant impact on the availabilities of the services, and hence is a crucial factor that should be taken into consideration for redundancy allocation.

\subsection{Comparison with Optimal Solution using ILP}\label{sec-6.2}  

\begin{figure*}[bth]
\centering
 \subcaptionbox{\textit{AllAny} model \label{utilinst}}{\includegraphics[scale=0.425]{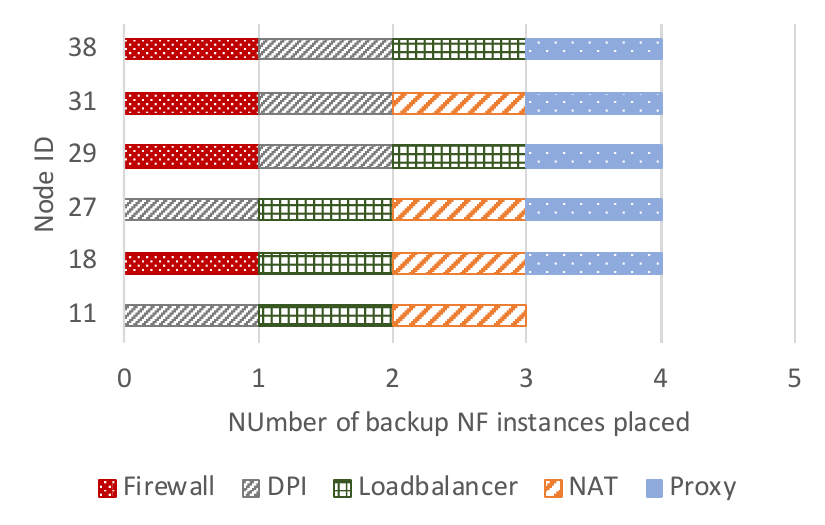}} 
  \hspace{0.25em}%
\subcaptionbox{CoShare: dedicated reservation \label{utilnode}}{\includegraphics[scale=0.425]{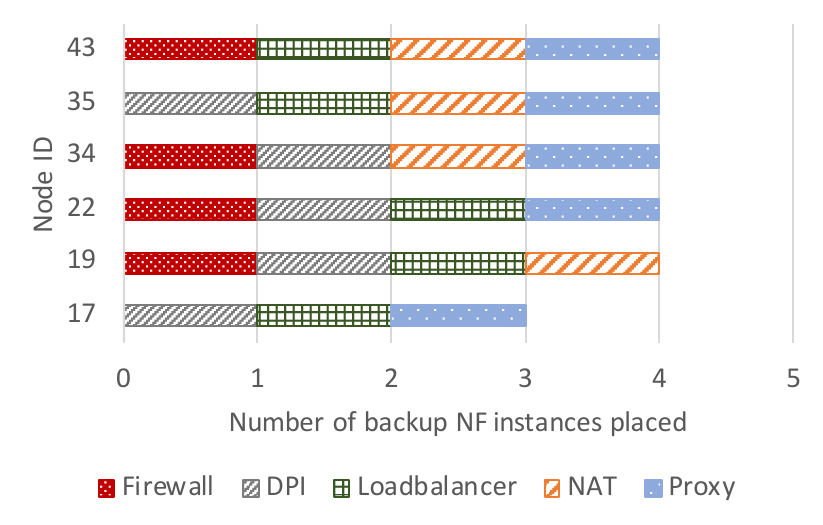}} %
\subcaptionbox{CoShare: shared reservation \label{utilnode}}{\includegraphics[scale=0.425]{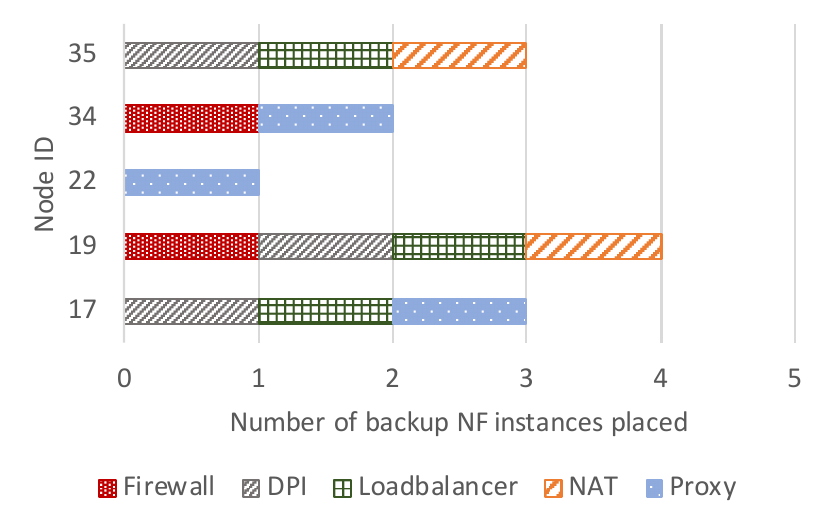}} %
 \caption{Placement of backup NF instances by the \textit{AllAny} model~\cite{woldeyohannes2019towards}, and CoShare using dedicated and shared reservations} \label{ilpcomplace}
\end{figure*}

\begin{figure*}[th]
\centering
 \subcaptionbox{\textit{AllAny} model \label{utilinst}}{\includegraphics[scale=0.425]{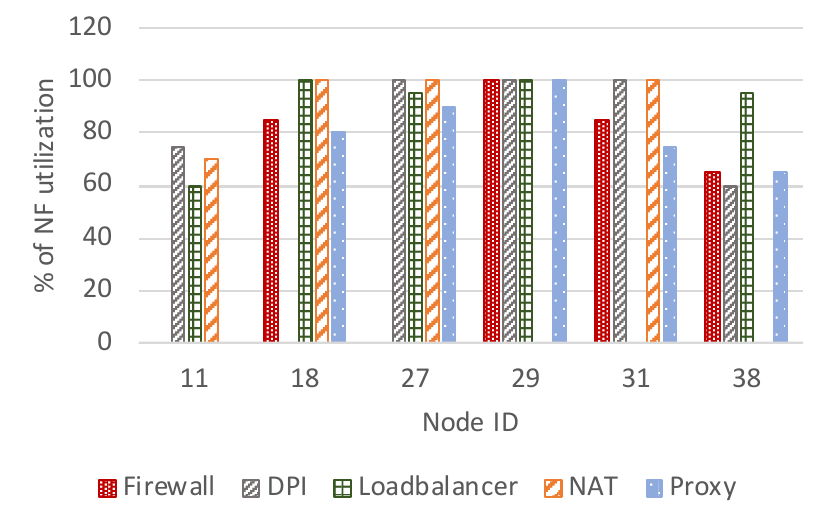}} 
  \hspace{0.25em}%
 \subcaptionbox{CoShare: dedicated reservation  \label{utilnode}}{\includegraphics[scale=0.425]{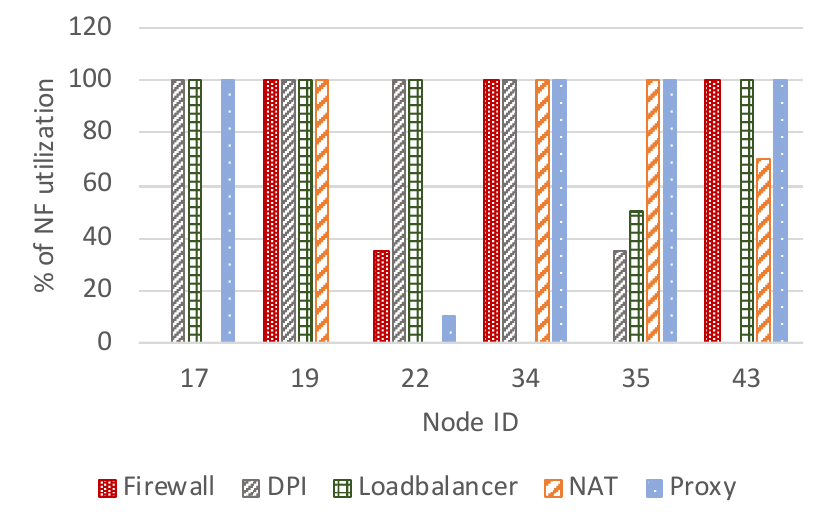}} %
  \subcaptionbox{CoShare: shared reservation \label{utilnode}}{\includegraphics[scale=0.425]{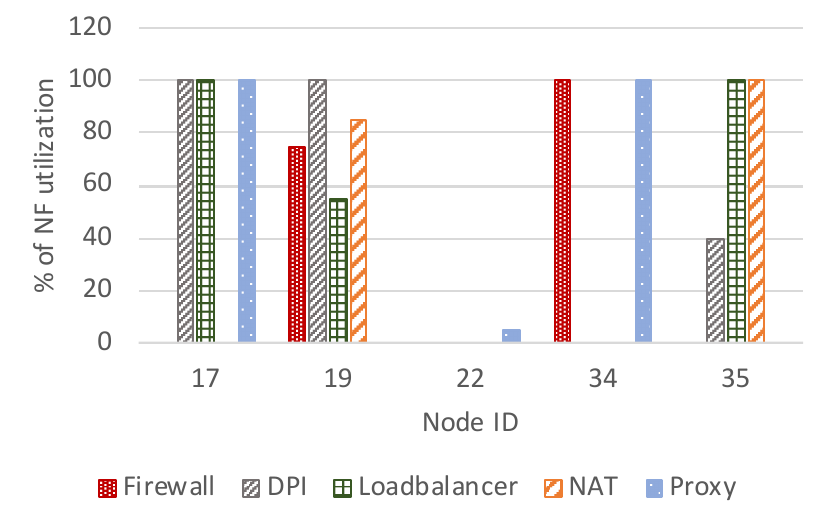}} 
 \caption{Per NF utilization of the placed backup NF instances} \label{ilpcomutl}
\end{figure*}

In order to assess the optimality level of CoShare, its performance is compared with an integer linear program (ILP) approach called \textit{AllAny}, which has been proposed  to solve the optimization problem (\ref{eq-obj}) in~\cite{woldeyohannes2019towards}. The ILP model assumes dedicated reserved capacity, i.e. dedicated reservation, at the backup instances.
The GEANT network is also used for this analysis. Backup chains are allocated for 200 flows whose primary chains are served by using 23 NF instances, where every flow has a chain length of 2, and its availability requirement is randomly set to be 3'9s, 4'9s, or 5'9s. 

For the comparison, we adopt the concept of \textit{resource overbuild}, which is a key figure-of-merit in assessing redundancy capacity efficiency~\cite{ou2004new},  
i.e. the extra capacity needed to meet the service availability objective as a percentage of the capacity for the service under no redundancy. In this paper, it is defined to be the ratio of the total number of backup NF instances actually used to meet the availability requirement to the total number of primary NF instances. 

Fig.~\ref{ilpcomplace} shows the placement of the backup instances, i.e., the number and type of backup NF instances placed together with their host nodes, obtained using the \textit{AllAny} model and CoShare. For CoShare, both shared reservation and dedicated reservation are considered. 

The number of backup instances for each NF which are created by CoShare with dedicated reservation is the same as that created by the \textit{AllAny} model. In total, 23 backup instances are created by each of the two approaches and six backup nodes are used to host the instances, i.e., 100\% resource overbuild. 
The per NF utilization level, which is the percentage of the NF capacity that is reserved by the flows, is shown in Fig.~\ref{ilpcomutl}. Since CoShare intends to maximize the utilization of NF instances, most of the backup instances are 100\% utilized in contrast to the NFs of the \textit{AllAny} model.
Moreover, with CoShare using shared reservation, only 13 backup NF instances are created and five backup host nodes are used, which results in only about 56\% resource overbuild. 

It is worth highlighting that in this example, the \textit{AllAny} model is solved by using the mathematical programming solver, CPLEX. However, even for this simple example, it took more than 10
minutes to find the optimal solution, due to the NP-hardness of the optimization problem (\ref{eq-obj}). In contrast, CoShare obtained the results in  one second.  
This showcases the fact that CoShare is able to get near optimal results but in much less  
time, and when applying shared reservations, CoShare can achieve better resource efficiency.

\begin{figure}[th!]
\centering
 \subcaptionbox{Number of backup NF instances \label{sharedinst}}{\includegraphics[scale=0.415]{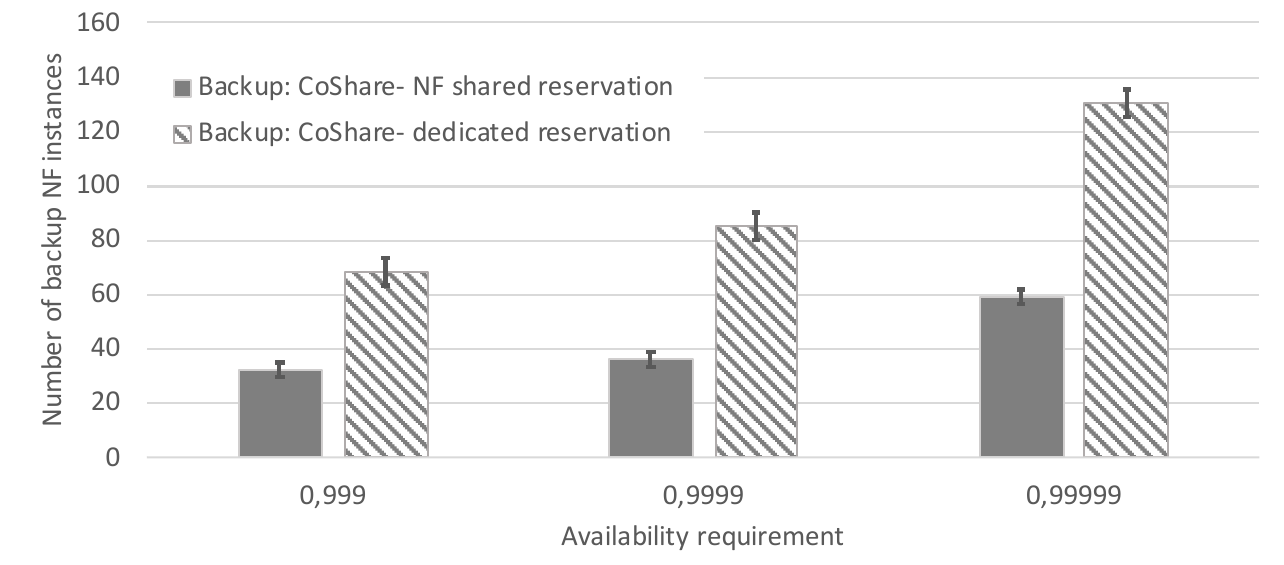}} 
  \hspace{0.25em}%
 \subcaptionbox{Resource overbuild \label{sharedover}}{\includegraphics[scale=0.52]{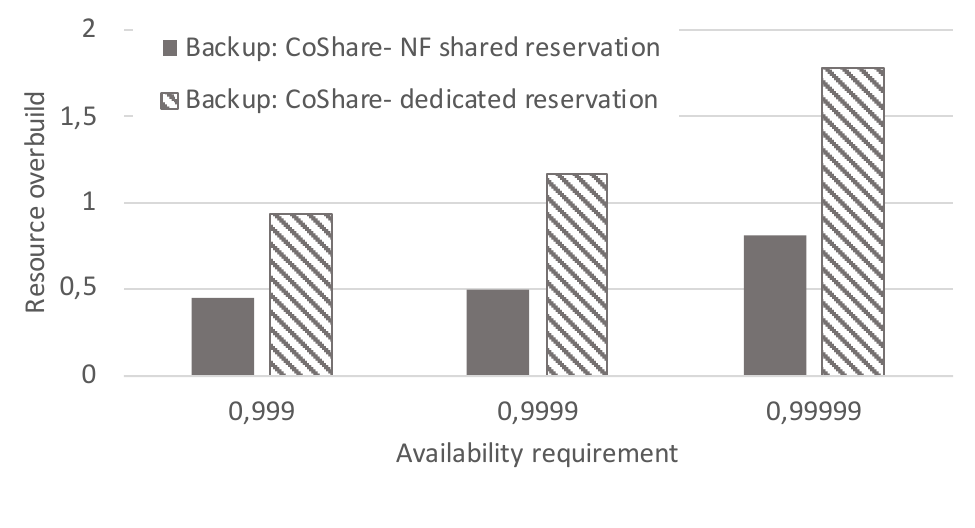}} 
  \caption{Comparison: Resource overbuild} \label{sharedreserv}
\end{figure}

\subsection{Effect of Shared Reservation} \label{sec-6.3}
We now consider a larger network: The Rocketfuel topology AS 1221 with 100 nodes and 294 links~\cite{spring2002measuring} is adopted. With the increased numbers of nodes and links, the computation of the optimal results has increased too much (due to NP-complexity) to be handled by the adopted workstation. For this reason, CoShare with dedicated reservation, which has similar performance as the optimal model shown in the above example, will be used in the comparison.

\subsubsection{Resource overbuild} \label{sec-bu-no} 
Fig.~\ref{sharedinst} shows the number of backup NF instances that are needed to satisfy different levels of availability by using CoShare with shared and dedicated reservation, where all flows require the same availability level. In each of the availability levels, the NF instances are created for 700 flows having a service chain containing two NFs. In addition, the resource overbuild for both backup allocation strategies is illustrated in Fig.~\ref{sharedover}. As can be observed, the higher the availability requirement level the more the number of backup NF instances required. This is because more than one backup chain might be required for fulfilling the high availability requirements. For example, more than 90\% of the flows with five nines availability requirements are allocated two backup chains.

\begin{figure}[th]
\centering
 \subcaptionbox{Number of backup NF instances\label{4inst}}{\includegraphics[scale=0.4]{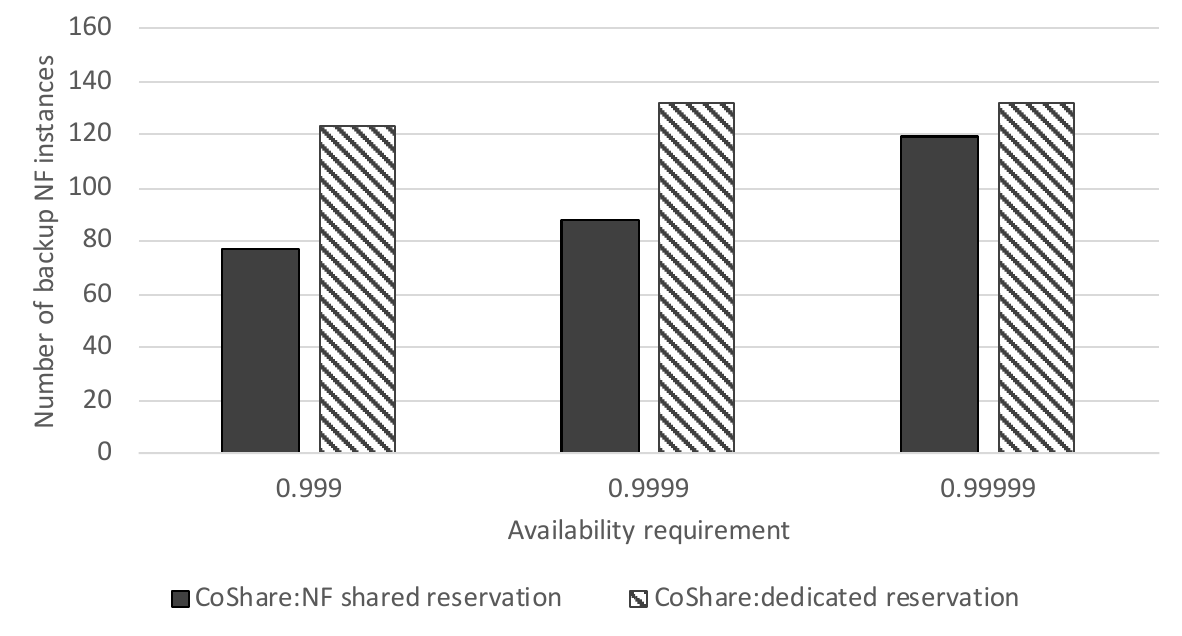}} 
  \hspace{0.25em}%
\subcaptionbox{Flow acceptance ratio \label{4flow}}{\includegraphics[scale=0.45]{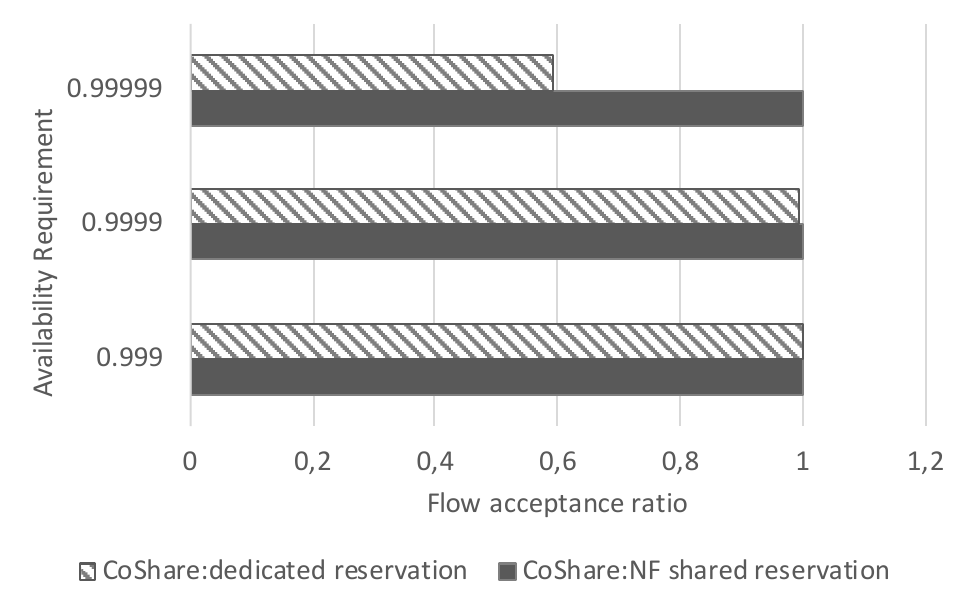}} 
 \caption{Comparison: Flow acceptance ratio} \label{chaineffect}
\end{figure}

\textit{Observation}: Recall that, in the placement phase, CoShare first ``roughly'' estimates the number of backup chains for each availability requirement class and accordingly the number of backup NF instances, i.e. $h_c$ and $z(c)$ in Sec. \ref{sec:nr_backups}, which are possibly needed. For the three availability levels [0.999, 0.9999, and 0.99999], the corresponding rough estimates are [1, 2, and 3]  for $h_c$ and [70, 140, and 210] for $z(c)$ respectively. As discussed in Sec. \ref{sec:nr_backups}, the actually used and hence needed numbers of backup instances under CoShare can be expected to be much lower. This is confirmed by Fig.~\ref{sharedinst} that shows the real total numbers of used backup instances under CoShare with dedicated reservation and shared reservation. Specifically under CoShare with dedicated reservation, they are [68, 85, 130] for the three availability levels, which are reductions of 2.8\%, 39\%, and 38\% from the rough estimates respectively. Under CoShare with shared reservation, there are further reductions of 44\%, 64\%, and 69\%.  

\textit{Finding}: Fig.~\ref{sharedover} further compares dedicated reservation and shared reservation using resource overbuild. As shown by the figure, to fulfill 0.99999 availability, the former requires 178\% resource overbuild, in contrast to the much reduced 93\% by the latter. Similar reduction is found also for the other two availability levels. Overall, shared reservation enables more efficient utilization of resources with significant decrease in the required number of backup NF instances.

\subsubsection{Flow acceptance ratio}
In the above experiments, no flow is rejected, i.e. all flows' availability requirements can be met with CoShare, with or without shared reservation. In the following experiment, we consider 650 flows each with a service chain consisting of four NFs. Similar to Fig.~\ref{sharedinst}, Fig.~\ref{4inst} compares the number of backup NF instances under dedicated and shared reservation. It also shows that less instances are needed with shared reservation. 

Additionally, Fig.~\ref{4flow} compares the flow acceptance ratio. As shown by the figure,  
all flows can be admitted with both shared and dedicated reservation when the availability requirements are under the two lower levels. 
However, when 5'9s availability level is required, only about $60\%$ of the flows can be admitted with dedicated reservation, in contrast to $100\%$ with shared reservation. This is because {\em CoShare with dedicated reservation} requires a higher number of backup NF instances than what can be provided by the network. If that number would have been possible, Fig.~\ref{4inst} would have shown an even higher reduction by using shared reservation. This again implies that NF shared reservation enables more resource efficiency which in turn maximizes the number of flows that can be admitted to the network.

\subsection{Effect of the Threshold}\label{sec-6.5} 
\label{sec:threshold}

\begin{figure}[ht!]
\centering
  \includegraphics[scale=0.40]{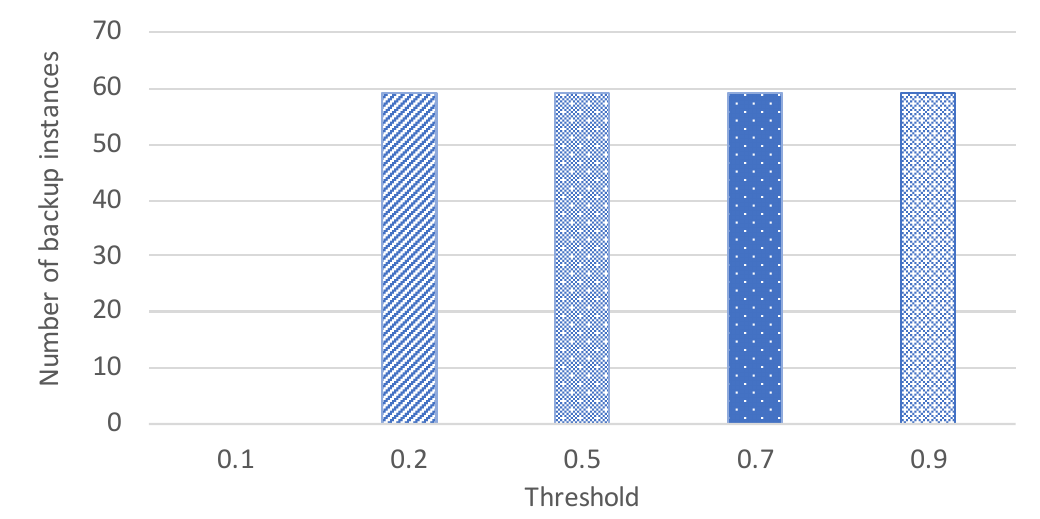} 
 \caption{Effect of $t_{DI}$} \label{EffectofThreshold}
\vspace{-10pt}
\end{figure}

CoShare has one algorithmic parameter, the threshold $t_{DI} \in (0,1)$, which is used in (\ref{e_critialset}) to help identify the set of critical nodes due to network structural correlation. As can be expected from (\ref{e_critialset}), a higher $t_{DI}$ leads to a smaller set.  To have a better overview about the effect of the threshold, experiments have also been conducted. Figure~\ref{EffectofThreshold} shows the number of backup NF instances instantiated for fulfilling the availability requirement of 0.99999 under different threshold values for the case of Rocketfuel topology with 700 flows. For the threshold value between 0.2 and 0.9, the same number is found. {\em With a closer look, it has been found that the same set of structurally correlated nodes are resulted 
from (\ref{e_critialset}) with a threshold value in this range, which constitutes the underlying reason for the effect shown by Figure~\ref{EffectofThreshold}.}
However, when $t_{DI} \in (0,1)$ is too small, e.g. 0.1, it is observed that all the network nodes are included in the set. These indicate that the performance of CoShare is generally robust to the threshold except when a too small value is given. 

\begin{figure*}[th]
\centering
 \subcaptionbox{Chain length=2 \label{proavail}}{\includegraphics[scale=0.35]{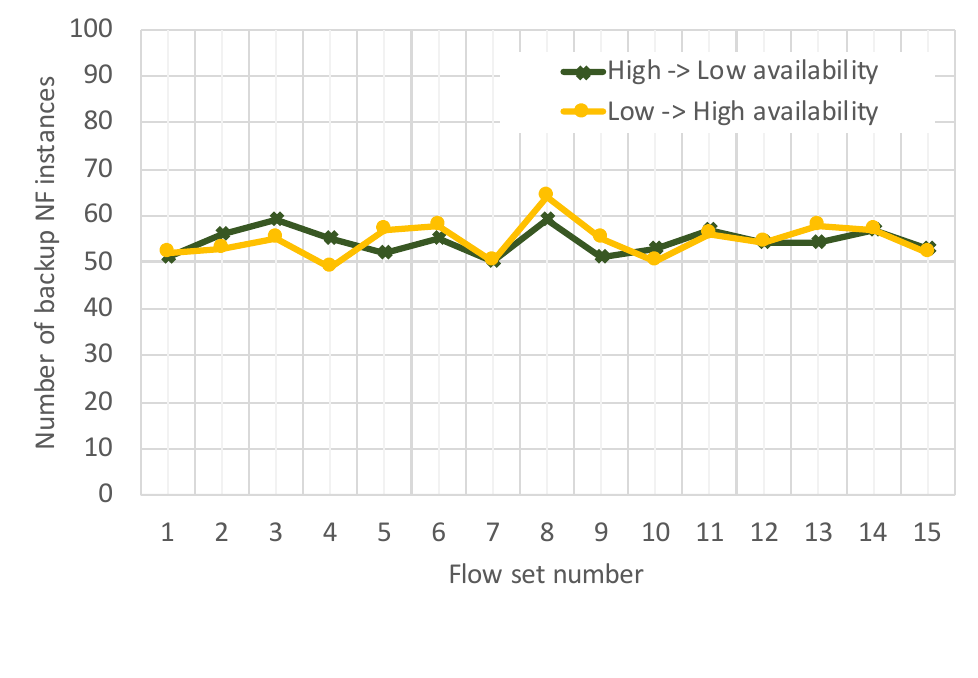}} 
  \hspace{0.25em}%
 \subcaptionbox{Low Availability class: 0.999 \label{procha3}}{\includegraphics[scale=0.35]{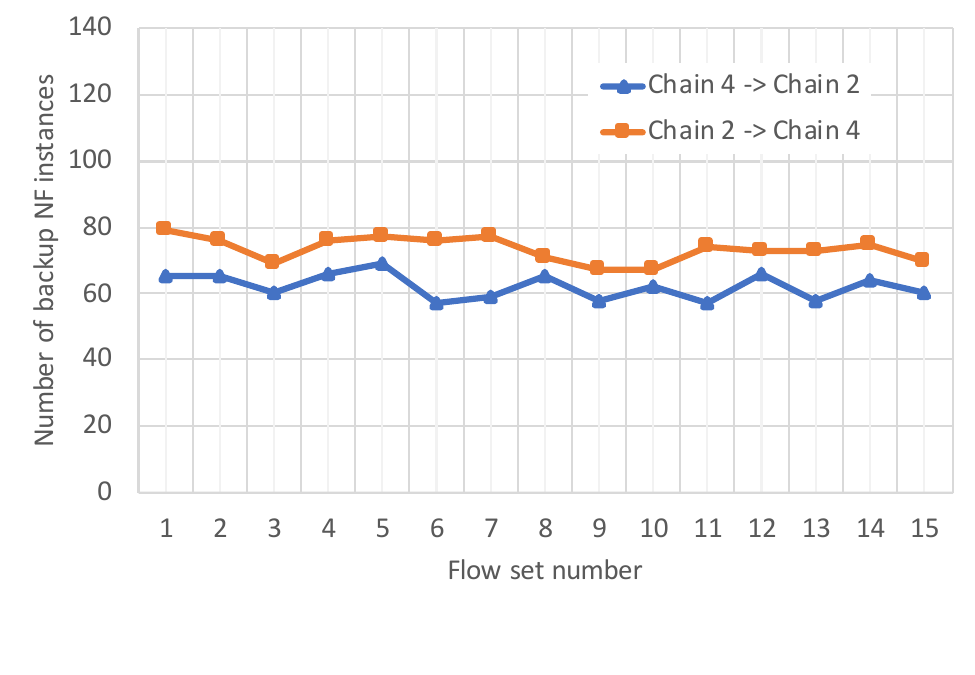}} %
  \subcaptionbox{High Availability class: 0.99999\label{procha5f}}{\includegraphics[scale=0.35]{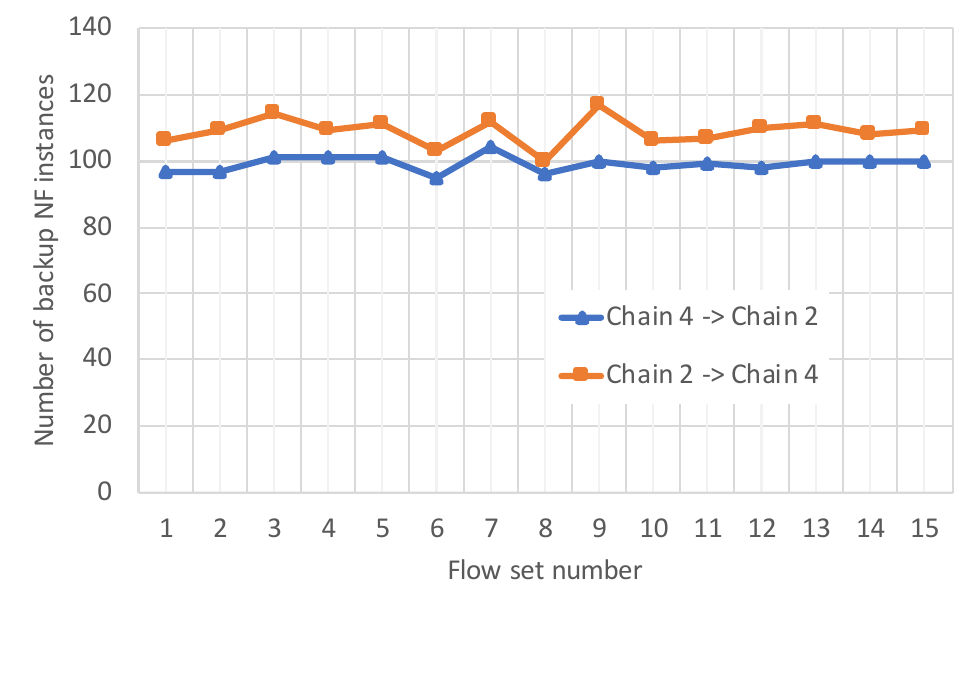}} %
 \caption{Effect of prioritization based on availability requirements (a) and service chain length (b) and (c).} \label{prioeffect}
\vspace{-10pt}
\end{figure*}

\subsection{Effect of Assignment Order}

Note that when CoShare is applied, flows are checked one-by-one in the step of assigning instances to form back chains for them (C.f. Sec.~\ref{assignment}). When there are multiple flows, these flows may be applied with CoShare at arbitrary order. In the rest, the effects of two specific yet intuitive orders are investigated.

In this investigation, the Rocketfuel topology is adopted. 15 sets of randomly generated 700 flows are considered. Each set is used for one simulation run. 
The availability requirement of each flow is randomly chosen among 3'9s, 4'9s and 5'9s. 
CoShare with shared reservation is focused. 

\subsubsection{Based on availability requirement}
 
First, we consider the effect of ordering the assignment based on the required availability levels. 
In this study, flows have the same service chain length of two. The flows are assigned with needed backup instance in the order of their availability requirements, from high (5'9s) to low (3'9s) or from low to high. 

Fig.~\ref{proavail} compares the total needed number of backup NF instances in fulfilling the availability requirements of flows in the 5'9s class under each simulation run using a different set of flows. As can be seen from the figure, there is no clear evidence about which order is better. Similar observation has also been found for the other two classes. This implies that prioritizing assignment based on availability requirements has minimal effect on the performance of CoShare.

\subsubsection{Based on service chain length}

Next, the impact of ordering based on the flow's service chain length is assessed: long to short or short to long. The experiment setup is similar to above, but the chain length of a flow is randomly chosen between 2 and 4. 

Figs.~\ref{procha3} and~\ref{procha5f} compare the number of needed backup NF instances for the 3'9s and 5'9s classes respectively. As can be observed from these figures, prioritizing flows that have longer service chain length results in fewer backup NF instances needed. One reason is that, when a longer chain flow is prioritized because of the involvement of more instances, the chance for a shorter flow that is assigned later finding and sharing capacity on some of these instances is higher. For example, in the best case, no new backup instance is needed. Nevertheless, the difference is minimal, about $10\%$ or less.

\subsection{Discussion}

The evaluation highlights the significant improvement that CoShare brings in terms of resource utilization efficiency and flow acceptance ratio. However, generalizing the results requires performance evaluation in more diversified NFV setups. We briefly present some aspects that could be further explored as future work. 

The placement decisions are subject to the underlying network topology structure and its intrinsic graph characteristics (i.e., node  degree, betweeness etc.). Aggregated results on a variety of network topologies, e.g., those present in SNDlib \cite{orlowski2010sndlib}, may provide more detailed insights about the effects of structural correlation on the placement and performance.  
In addition, it is common that NFV-enabled network operators provide services through well-defined service chains, typically included in their NFV service catalogs, rather than through a random composition of NF instances. An alternative evaluation would be to consider a set of specific service chains and characterize their availability demands and probability of occurrence, i.e., probability of being requested.
Moreover, we made some simplifying assumptions compared to a more truthful representation of network nodes' and links' capacities. In realistic networks, some of the nodes could be central offices with rather limited capabilities, whereas some other nodes could be data centers with significantly more computing resources. 
 Besides, we have implicitly assumed that all links have the needed capacity. However, failure events may trigger traffic re-routing, increasing the utilization levels of certain links. A related aspect concerns cases where multiple distributed copies of an NF on the service chain may be needed to work together, to deliver the service, e.g., a load-balancer. 
To address these needs, CoShare may be extended to include link utilization constraints and to consider more generalized ways of defining service chains. Clearly, there will be an increase in the problem complexity, as well as a modification of the assignment heuristics to account for the link utilization constraints and to support multiple copies of one NF on a chain. This could be another direction for future  extension of CoShare.

\section{Conclusion}\label{sec-conc}
In this paper, a novel scheme, called CoShare, is proposed for redundancy allocation in NFV. An original and crucial idea of CoShare is to explicitly take into account the inherent network structure-caused correlation / dependence among nodes 
in both redundancy placement and assignment. As a result, CoShare is able to minimize the impact of correlated failures due to network structural dependence on service availability. 
In addition, CoShare allows {\em shared reservation} among flows, i.e. let them share the same reserved backup capacity at an instance, to improve resource efficiency without compromising their availability. This forms another contribution of CoShare. 
Moreover, CoShare stands out with supporting diverse flow availability requirements under heterogeneous nodes and instances in terms of both resources and availability.   
The experimental results demonstrate that a redundancy approach without considering network structural dependence in its design can unfortunately fail to meet its promised availability. In addition, when backup capacity is dedicatedly reserved for each flow at a shared NF instance, the performance of CoShare (with dedicated reservation) is close to that of the optimal solution,  but CoShare is scalable. Furthermore, with shared reservation, CoShare can reduce the resource overbuild significantly, e.g. about half or more in the Rocketfuel topology experiment. These results indicate that CoShare is appealing for redundancy allocation in NFV. They also imply the criticality and potential of taking into account network structural dependence in addressing the NFV redundancy allocation problem.

\bibliographystyle{IEEEtran}
\bibliography{reference-compact}
\newpage
\appendices
\section{}

\begin{IEEEproof}[Proof of Theorem 1]\label{ap1}
 Suppose there is one NF instance on which flow $f$ can share reserved capacity in the backup chain $r$. Then, the chain weight will be 

\begin{equation}
W^p(r,f)=g_f +  \sum_{g=1}^{g_f - 1} \frac{\sum_{f_a \in \mathcal{D}_{v^r_{f,g}}}\lambda_{f_a}}{\mu_{v^r_{f,g}}}.
\end{equation}
For the chain $r'$, it has no NF instance on which the flow can share reserved capacity, i.e., flow $f$ is not independent with all the flows in $\mathcal{D}_{v^{r'}_{f,g}}$, $\forall g \in \{1 \ldots g_f \}$. Then, the chain $r'$ will have a weight 
\begin{equation}
W^p(r^{'},f)=  \sum_{g=1}^{g_f} \frac{\sum_{f_a \in \mathcal{D}_{v^{r^{'}}_{f,g}}}\lambda_{f_a}}{\mu_{v^{r^{'}}_{f,g}}}.
\end{equation}

Note that, the NF instances on all the backup chains in $\mathcal{R}(f)$ satisfy the NF capacity constraint. Thus,

\begin{equation}
0 \leq \frac{\sum_{f_a \in \mathcal{D}_{v^{r^{'}}_{f,g}}}\lambda_{f_a}}{\mu_{v^{r^{'}}_{f,g}}} < 1 \; \text{ and, } \; 0 \leq \frac{\sum_{f_a \in \mathcal{D}_{v^r_{f,g}}}\lambda_{f_a}}{\mu_{v^r_{f,g}}} < 1.
\end{equation}
Since the service chain length $g_f \geq 1$ and the other conditions are the same, we have $W^p(r,f) > W^p(r^{'},f) $. When there are more instances on $r$ where the flow can share reserved backup capacity, following the same derivation as above, it can be easily verified that the weight of $r$ will then be even higher. This concludes the proof.
 \end{IEEEproof}
 
\begin{IEEEproof}[Proof of Theorem 2]\label{ap2}

The complexity of CoShare is determined by the three involved parts, namely network structural analysis (Sec. \ref{sec:dependency_index}), placement (Sec. \ref{sec:heuristic}), and  assignment (Sec. \ref{assignment}). In network structural analysis, the computational complexity of finding the critical node set $\mathcal{C}(n)$ for every node $n \in \mathcal{N}$ is $O(N^2)$ from (\ref{e_critialset}) and (\ref{nodedpl}), so its complexity is $O(N^3)$. CoShare's placement heuristic performs a bin-packing of backup NF instances on nodes. The complexity of the algorithm is a function of the number of instances to be placed and the number of candidate backup host nodes. Specifically, sorting is performed on both the node side and the instance side, whose complexities are $O(N^2)$ are  $O(z_v^2)$ respectively, where $z_v$ is the maximum number of backup instances (\ref{eq-est}) that may be involved in the sorting. Approximating the number of NF instances by the number of nodes, the placement heuristic has a complexity $O(N^2)$. The assignment of NF instances to form backup chains for flows has a complexity of $O(FN^G)$, where $G$ denotes the longest service chain length. This is because, for every flow, the worst case is to search through all the possible chain compositions and the total number is $|\mathcal{I}_{S_f^1}| \cdots |\mathcal{I}_{S_f^{g_f}}|$ which is upper-bounded by $O(N^G)$. Thus, the complexity of CoShare is $O(N^3 + N^2 + FN^G)$, which is approximately $O(FN^G)$, under practical assumptions $F \ge N$ and $G \ge 2$.
\end{IEEEproof}

\section{}
The list of notations used in problem formulation is shown in Table \ref{table_e}.

\begin{table}[h!]
\footnotesize
\caption{List of notations used in problem formulation.}
\label{table_e}
\centering
\resizebox{\columnwidth}{!}{
\begin{tabular}{l|l}
\hline
\bfseries Notation & \bfseries Definition\\
\hline\hline
Network Model & \\ \hline\hline
$\mathcal{N}$ & set of nodes\\
$\mathcal{L}$ & set of links\\
$\mathcal{F}$ & set of flows requiring service\\
$\mathcal{V}$ & set of NF instances\\
$\mathcal{N}_{b_f}$ & set of nodes hosting backup chain instances of $f$\\
$\overrightarrow{S}_{f}$ & service chain of flow $f$, $f\in \mathcal{F}$\\
$g_f=|\overrightarrow{S}_f|$ & service chain length of flow $f$\\ 
$k_v$ & capacity (CPU cores) request of NF $v$, $v\in \mathcal{V}$\\ 
$\lambda_f$ & flow's $f$ traffic rate\\ 
$\mu_v$ & processing capacity of NF $v$\\
$p_f$ & primary chain of flow $f$ \\
$b_f$ & backup chain of flow $f$ \\
$\mathcal{F}_v$ & number of flows processed by NF $v$\\ \hline\hline

Availability Model & \\ \hline\hline
$A_{n}$ & availability of node $n$, $n\in \mathcal{N}$\\ 
$A_{v}$ & availability of NF $v$\\ 
$A_{r_f}$ & availability requirement of flow $f$\\ 
$A_{p_f}$ & availability of the primary chain of flow $f$\\ 
$A_{b_f}$ & availability of the backup chain $b_f$ of flow $f$, $b_f \in B_f$\\ 
$v_g$ & $g$-th NF instance of backup chain $b_f$\\ 
$h_f=|B_f|$ & number of backup chains for flow $f$\\ \hline\hline

Structural Dependence & \\ \hline\hline
$DI(i|n)$ &  node dependency index, $i,n\in \mathcal{N}$\\ 
$I_{ij}$ &  information measure among \\
& nodes $i$ and $j$, $j\in \mathcal{N}$\\ 
$I_{ij}^{-n}$  & information measure among \\
& nodes $i$ and $j$ after removing node $n$\\ 
$d_{ij}$ & length of shortest path among nodes $i$ and $j$\\ 
$d_{ij}^{-n}$ & length of shortest path among nodes $i$ and $j$ after\\
& removal of node $n$\\ 
$\mathcal{A}_{ij}^{-n}$ & reachability of node $j$ from node $i$ given that\\
& node $n$ is removed\\ 
$\mathcal{N}^{-n}$  & set of nodes in the network after removal of node $n$\\ 
$\mathcal{C}(i)$  & set of critical nodes of node $i$\\ 
$\hat{\mathcal{B}}_i$ & set of nodes that are network-structurally \\
& correlated with node $i$\\\hline \hline

Placement strategy & \\ \hline\hline
$\mathcal{F}_{c}$ & set of flows belonging to class $c$, $c\in \mathcal{C}$\\ 
$h_c$ & number of backup chains needed to satisfy \\
& the requirement $\forall f\in \mathcal{F}_{c}$\\ 
$A_{c}^{r}$ &  maximum availability requirement\\
& of flows for class $c$\\ 
$z_{v}(c)$ & number of backup instances of NF $v$ \\
& for flows of class $c$\\ 
$z_{v}$ &  total estimated number of backup instances of NF $v$\\ \hline\hline

Assignment strategy & \\ \hline\hline
$\tilde{\mathcal{I}}_{S_f^{g}}$ & set of feasible NF instances to act\\
& as backup for the $g$-th NF of flow $f$\\ 
$\mathcal{R}(f)$ & set of feasible backup chains for flow $f$\\ 
$w(v^r_{f,g})$ & weight of the instance of the $g$-th NF \\
& in chain $r$ of flow $f$\\ 
$W(r,f)$ & sum of the weight of each instance composing chain $r$\\ 
$\mathcal{SR}_{v}(f)$ & set of flows sharing instance's $v$ capacity with flow $f$ \\ \hline\hline
\end{tabular}
}
\end{table}

\section{Related Work}
\label{sec-rela}
\begin{table*}[t!]
\caption{Comparison of previous works on NFV availability-guaranteed resource allocation (RA).}
\label{table_example}
\centering
\resizebox{2\columnwidth}{!}{
\begin{tabular}{|ccccccc|}
\hline
\bfseries Ref & \bfseries Objective & \bfseries NFV RA Problem & \bfseries Method & \bfseries Protection & \bfseries Types of failure & \bfseries Correlated failures\\ \hline\hline
[8] & max. \#SFCs & Placement & INLP \& Heuristic & Dedicated & VNFs & No \\  \hline
[13] & min. \#Nodes & Placement \& Assignment & ILP & Dedicated & Links \& Nodes & No\\  \hline
[15] & min. \#(VNFs$+$Nodes) & Placement  & INLP \& Heuristic & Dedicated  & VNFs \& Node & No\\  \hline
[16] & max. VNF availability & Placement & Bipartite-graph assignment \& Heuristics & Dedicated & VNFs$^1$ & No\\  \hline
[17] & min. VNF Unavailability & Placement & MILP & Dedicated & VNFs & No\\  \hline
[18] & min. \#VNFs & Placement & Heuristic & Joint$^2$ & Nodes & No\\ \hline
[19] & min. Link BW & Assignment & ILP & Dedicated \& Shared$^3$ & Links & No\\ \hline
[20] & min. Link BW & Placement \& Assignment & ILP \& Heuristic & Shared$^4$ & Nodes & No\\ \hline
[21] & min. \#Nodes & Placement & Heuristic & Shared$^4$& VNFs & No\\ \hline
[31] & min. \#Resources & Placement & INLP \& Heuristics & Dedicated \& Shared$^4$& VNFs & No\\ \hline
[36] & min. \#Nodes & Placement \& Assignment & Heuristic & Dedicated & Nodes & No\\ \hline
[43] & max. SFC Reliability & Placement & Heuristic & Dedicated & Nodes & No\\ \hline
[46] & max. SFC Availability & Placement & MILP \& Heuristic & Dedicated & VNFs \& Nodes & No\\ \hline
[47] & max. SFC Availability & Placement & Bipartite-graph assignment \& Heuristics & Dedicated & VNFs$^1$ & No\\ \hline
[48] & max. \#SFCs & Placement \& Assignment & Heuristic & Dedicated & -- & No\\ \hline
This work & min. \#(VNFs$+$Nodes) & Placement \& Assignment & Heuristic & Dedicated \& Shared & VNFs \& Nodes & Yes\\ \hline
\end{tabular}
}
{\raggedright $^1$\textit{Only primary VNFs} \par}
{\raggedright $^2$\textit{A combination of dedicated and intra-SFC sharing but with resources equal to a dedicated version.} \par}
{\raggedright $^3$\textit{Link bandwidth sharing} \par}
{\raggedright $^4$\textit{Intra-SFC sharing between two or more adjacent VNFs} \par}
\end{table*}


Guaranteeing service availability in NFV-enabled networks represents an important challenge that needs to be addressed to fully exploit the benefits of NFV~\cite{mijumbi2016nfv_challenges,tola2019network}. To this aim, there has been a continuous effort in the recent literature to investigate and propose resource efficient and scalable algorithms  for resource / redundancy allocation in NFV. 

Fan \textit{et al.} \cite{fan2015grep} presented an algorithm to minimize the employed physical resources by protecting the most unreliable NFs. On similar lines, they extended the work by proposing methods for allocating backup resources in order to maximize the number of accommodated service requests while meeting heterogeneous availability demands \cite{fan2017carrier}. In \cite{herker2015data}, the authors studied the suitability of various data center architectures for resilient NFV service deployments. Ding \textit{et al.} \cite{ding2017enhancing} improved the design in \cite{fan2015grep} by proposing a method to select the most appropriate NFs to protect by exploiting a cost-aware critical importance measure rather than the least available NFs. The work in \cite{kanizo2017optimizing} investigates the problem of backup instance placement, called assignment in the work, to hosting nodes for maximizing the probability for a full recovery. The same authors design recovery schemes in a multi-failure scenario while guaranteeing full recovery of failing instances based on a novel graph-based approach~\cite{kanizo2018designing}.
However, these contributions are based on assumptions which may significantly impact their applicability in more general setups~\cite{zhang2019raba,herker2015data}. Such assumptions include homogeneous backup nodes, considering only the failure of NFs while ignoring physical nodes' failure and vice versa \cite{fan2015grep,ding2017enhancing}, assuming NF instances fail independently irrespective of their placement~\cite{fan2017carrier}, or even assuming that only primary instances can fail as in \cite{kanizo2017optimizing,kanizo2018designing}. Moreover, these contributions address only the placement sub-problem of the more complex NFV redundancy allocation problem~\cite{fan2017carrier,kanizo2017optimizing,fan2015grep,herker2015data,ding2017enhancing,kanizo2018designing}.

In \cite{hmaity2017protection}, three ILP models are proposed for VNF placement and service chaining. However, their aim is to protect the service chains against 
different types of failure 
without taking into account the specific availability requirement of each service chain. In addition, the evaluation shows that providing protection against the considered failure scenarios comes with at least twice the amount of resources in terms of the number of nodes being deployed into the network. Similarly, \cite{yang2018algorithms} optimizes the backup NF placement and chaining with the objective of maximizing the number of admitted service requests. Owing to the NP-hardness of the problem, they propose a heuristic algorithm for maximizing data-center resource utilization. However, the work limits the fault-tolerance provisioning to only placing a standby instance without requiring the fulfillment of specific service availability demands.

As redundancy can be costly, it is desirable to share redundancy at maximum possible in NFV based networks to enable more efficient resource utilization as having been done in traditional networks,  e.g., \cite{ramamurthy1999survivable,li2003efficient}. 
In~\cite{li2019availability}, a multi-tenancy based approach, which allows a backup NF instance to be utilized by multiple flows, is proposed, and it is also demonstrated that the multi-tenancy based approach outperforms single-tenancy based approaches. However, in the approach proposed in~\cite{li2019availability}, a backup chain is constrained to only be using NFs hosted on one node, similar to~\cite{fan2018framework}.  
In \cite{fan2015grep}, aiming at minimizing the physical resource consumption, a joint protection scheme is proposed where the sum of resources between two adjacent NFs are allocated for protection. In~\cite{tomassilli2018resource}, shared path protection is used to allocate backup paths that protect against single link failures. In~\cite{qu2018reliability}, adjacent NFs share the resources of a host machine. 
In all these approaches, when a backup NF instance is assigned to a flow, dedicated backup capacity for the flow is reserved, same as in CoShare with {\em dedicated reservation}. 

In~\cite{CASAZZA201947}, the authors propose a flexible VNF placement, which enables VNF protection only if needed, and the redundant VNF can also be shared among multiple active instances. However, the work performs only VNF placement and does not address the flow assignment problem. Moreover, both \cite{CASAZZA201947} and \cite{he2019optimization}, although they acknowledge the potential impact of common cause failures, i.e., correlated, and indicate possible extensions of their protection mechanisms to cope with these failures, they do not investigate the impact of such failures.

To summarize, Table II presents the references investigating the NFV backup resource allocation problem and highlights key characteristics of the applied methodologies.

CoShare is different from these literature works in several aspects. First, CoShare takes into account the heterogeneity present in terms of resources at network nodes and NF instances and their availability, in contrast to~\cite{zhang2019raba,herker2015data}. Second, both node failure and NF failure as well as the impact of a node's failure on its hosted NFs are considered, different from \cite{fan2015grep,ding2017enhancing} and \cite{fan2017carrier}. Third, CoShare aims to meet flow-level specific availability requirement, different from \cite{hmaity2017protection}. 

Fourth, in terms of shared reservation, to enable resource efficiency, CoShare allows a backup service chain being constructed by NF instances placed at different nodes, as opposed to the approaches studied in \cite{li2019availability, fan2018framework}. In addition, CoShare also differs from \cite{fan2015grep,qu2018reliability}. Specifically, the shared reservation mechanism employed in CoShare provides protection to multiple service chains that request the same NFs, rather than protecting adjacent NFs of the same service chain in \cite{fan2015grep,qu2018reliability}. It is worth highlighting that in all the literature approaches \cite{fan2015grep,qu2018reliability,li2019availability, fan2018framework}, even though  a backup NF instance may be shared among multiple flows or tenants, dedicated capacity is reserved for each flow / tenant, the same as in {\em dedicated reservation} discussed in Sec. \ref{sec:efficient}. In other words, as in CoShare's dedicated reservation, resource sharing in all these approaches is at the instance level. However, CoShare's shared reservation {\em additionally} allows the sharing to be made at the flow or service chain level, leading to improved efficiency in making use of resources. 
Moreover, none of the previous works takes into consideration topological dependencies among network nodes. Since such dependencies are inherent in the network structure / topology, disregarding them could lead to the failure of both the primary and the backup chains at the same time, and consequently affects the actually delivered availabilities of flows lower than expected as shown by the example in Sec.~\ref{sec-6.1}. To this end, CoShare not only makes another novel contribution but also sheds a new insight for redundancy allocation in NFV. 
\end{document}